\documentclass[oldversion]{aa}

\usepackage{txfonts}
\usepackage[]{graphicx}
\usepackage[]{epsfig}
\usepackage[]{longtable}
\usepackage[]{hhline}
\include{epsf}
\usepackage{natbib}
\begin{document}

\def\fdeg{\hbox{$.\mkern-4mu^\circ$}}
\def\farcs{\hbox{$.\!\!^{\prime\prime}$}}
\def\farcm{\hbox{$.\mkern-4mu^\prime$}}
\def\degr{\hbox{$^\circ$}}
\def\arcmin{\hbox{$^\prime$}}
\def\arcsec{\hbox{$^{\prime\prime}$}}
\def\sun{\hbox{$_\odot$}}
\def\logg{\hbox{$\log g$}}
\def\Teff{\hbox{$T_{\rm eff}$}}
\def\fh{\hbox{$.\!\!^{\rm h}$}}
\def\fm{\hbox{$.\!\!^{\rm m}$}}
\def\fs{\hbox{$.\!\!^{\rm s}$}}
\def\hou{^{\rm h}}
\def\min{^{\rm m}}
\def\ssec{^{\rm s}}

\def\logg{log\hspace*{1mm}$g$}
\def\mfe{$\langle{\rm Fe}\rangle$}
\def\mh{$\langle{\rm H}\rangle$}
\def\mv{$M_V$}
\def\dcn{${\rm \delta CN}$}

\def\unit#1{\;{\rm #1}}


\def\wcen{$\omega$\,Cen}
\def\wcentauri{$\omega$\,Centauri}

\def\apj{ApJ}
\def\aap{A\&A}
\def\aaps{A\&AS}
\def\araa {ARA\&A}
\def\aj{AJ}
\def\apjl{ApJ}
\def\apjs{ApJS}
\def\asp{ASP}
\def\pasp{PASP}
\def\physrep{Phys.Rep.}
\def\apss{Ap\&SS}
\def\nat{Nature}
\def\mnras{MNRAS}
\def\pasj{PASJ}


\title{Comparing CN and CH Line Strengths in a Homogeneous Spectroscopic 
Sample of 8 Galactic Globular Clusters\\ 
\thanks{Based on observations obtained at the European Southern Observatory, 
Chile (Observing Programmes 69.D-0172 and 73.D--0273).}}

\author {Andrea Kayser \inst{1,2} \and Michael Hilker \inst{3,4} \and Eva K. 
Grebel \inst{1,2} \and Philip G. Willemsen \inst{3}}

\offprints {A.~Kayser, \email{Andrea.Kayser@unibas.ch}}

\institute{
Astronomisches Institut der Universit{\"a}t Basel, Venusstrasse 7, 4102 Binningen, Switzerland
\and
Astronomisches Rechen-Institut, Zentrum f{\"u}r Astronomie, Universit{\"a}t Heidelberg, M{\"o}nchhofstra{\ss}e 12-14, 69120 Heidelberg, Germany
\and
Argelander-Institut f{\"u}r Astronomie, Auf dem H{\"u}gel 71, 53121 Bonn, Germany
\and
ESO, Karl-Schwarzschild-Str. 2, 85748 Garching bei M{\"u}nchen, Germany
}

\date{Received / Accepted }

\titlerunning{CN and CH line strengths in MW GCs}

\authorrunning{A.~Kayser et al.}

\abstract{Our work focuses on the understanding of the origin of CNO-anomalies, which 
have been detected in several Galactic globular clusters.
The novelty and advantage of this study is that it is based on a homogeneous 
data set of hundreds of medium resolution spectra of stars in eight Galactic 
globular clusters (M\,15, M\,22, M\,55, NGC\,288, NGC\,362, NGC\,5286, 
Palomar\,12 and Terzan\,7). Two of the clusters (Palomar\,12 and Terzan\,7) 
are believed to be former members of the Sagittarius dwarf spheroidal (Sgr 
dSph) galaxy. The large homogeneous data set allows for a detailed 
differential study of the line strengths in the stellar spectra of the 
observed clusters. Our sample comprises stars in different evolutionary 
states, namely the main-sequence turn-off (MSTO) region, the subgiant branch 
(SGB) and the base of the red giant branch (RGB). We compare the relative CN 
and CH line strengths of stars in the same evolutionary states (with similar 
\logg\ and \Teff ). The majority of the examined clusters show significant 
variations in their CN and CH abundances at the base of the RGB. We confirm 
the presence of a bimodal distribution in CN for the second parameter pair of the clusters 
(NGC\,288 and NGC\,362). The two probable former Sgr dSph clusters do not 
exhibit any CN-strong stars. Overall, our results suggest that the environment in which the 
clusters formed is responsible for the existence of CN-strong stars.
We can confirm the known anticorrelation between CN and CH for most of the 
observed clusters. Although the signal of CN absorption is weaker for the 
hotter stars on the MSTO and SGB we observed the same anticorrelation in these
less evolved stars for the CN-bimodal clusters.
Including structural parameters taken from literature reveals
that the existence of the CN-bifurcation seems to be independent from most other
cluster characteristics. In particular, we do not confirm the correlation 
between cluster ellipticity and number of CN-strong stars.
However, there may be a trend of an increased percentage of CN-strong stars 
with increasing cluster tidal radius and total luminosity.
We argue that our findings are consistent with
pollution by intermediate AGB stars and/or fast rotating massive stars
and two generations of star formation in luminous clusters with larger tidal radii
at larger Galactocentric distances.}


\maketitle


\section{Introduction}\label{intro}
Among the about 150 known Galactic globular clusters (GC) there exist several 
clusters that show star-to-star abundance variations for certain chemical 
elements \citep[see review from][]{G/S/C:04}. These variations are ubiquitous 
particularly for light elements such as C and N and are seen mainly for stars 
on the red giant branch (RGB). Stars with significantly stronger cyanogen (CN) bands
as compared to other stars in the same cluster have been detected as early as 
1971 by \citeauthor{O:71} in M5 and M10 \citep{O:71}. The existence of such 
stars in these and many other clusters has been confirmed repeatedly 
\citep[e.g.,][]{C:78, S/N:82, S/N:83, B:89}. However, the fraction of red 
giants showing enriched CN bands differs from cluster to cluster \citep{N:87}.

\begin{table*}[t!]
\centering
\caption{\label{tab1}Log of observations}
\begin{tabular}[l]{clrrr}
\hline\hline
 Date & Target & \multicolumn{2}{c}{RA; DEC (J2000)} & Exp.time \\
 \hline
 May 2002 & M55 MSTO 		& 294.99564 & -30.88307 &  $1800 \unit{s}$ \\
	  	& M55  SGB 		& 294.99646 & -30.88368 & $2160 \unit{s}$ \\
	  	& M55  RGB 		& 294.99559 & -30.88235 & $ 480 \unit{s}$ \\
  \hline  
 July 2004 & NGC\,288 MSTO 	& 13.23313 &  -26.57845 &  $ 5140 \unit{s}$ \\
	   	& NGC\,288  SGB 	& 13.23630 &  -26.57807 & $ 2700 \unit{s}$ \\ \hhline{~----}
 	   	& NGC\,362 MSTO 	& 15.67363 &  -70.84870 & $ 5400 \unit{s}$ \\
           	& NGC\,362  SGB 	& 15.67209 &  -70.84886 &  $ 2800 \unit{s}$ \\ \hhline{~----}
          	& NGC\,5286 SGB 	& 206.54375 &  -51.37364 & $ 2700 \unit{s}$ \\ \hhline{~----}
	   	& M\,22      MS 		& 279.04539 &  -23.90313 &  $ 5400 \unit{s}$ \\
	   	& M\,22     SGB 		& 279.04539 &  -23.90311 &  $ 3000 \unit{s}$ \\ \hhline{~----}
	   	& Ter\,7    SGB 		& 289.43484 &  -34.65680 &  $ 5400 \unit{s}$ \\
	   	& Ter\,7    RGB 		& 289.43488 &  -34.65773 &  $ 4500 \unit{s}$ \\ \hhline{~----}
	   	& M\,15     SGB 		& 322.54426 &   12.16722 &  $ 2400 \unit{s}$ \\ \hhline{~----}
	   	& Pal\,12   RGB 	& 326.66087 &  -21.25134 &  $ 2400 \unit{s}$ \\ 
\hline\hline
\end{tabular}
\end{table*}

Over the last three decades spectroscopic studies of the CN and CH absorption 
bands often revealed a bimodality in CN that is accompanied by a broadened 
distribution in CH. For the majority of the CN-bimodal clusters (e.g., M\,2, 
M\,3, M\,5, M\,13, 47\,Tuc) a CN-CH anticorrelation was detected 
\citep[e.g.,][]{S/S:96}. Since CN is a double-metal molecule, it can
be more easily observed in stars with a higher metallicity. Nevertheless, the
CN-CH anticorrelation seems to be present also in the very metal-poor cluster
M15 where no clear bimodality of CN could be detected so far \citep[][]{L:00}.

Although this topic has been studied extensively in the last decades no 
self-consistent model has been found to satisfactorily explain the observed 
chemical variations. Two main scenarios are discussed as possible
origins of these abundance patterns:

1) The `evolutionary mixing' scenario: In this scenario the chemical 
composition in the surfaces of the stars is altered due to deep mixing effects.
Material from the stellar interior is dredged-up through regions of active 
CNO element nucleosynthesis to the upper layers of H-burning. During the 
H-burning phase via the CNO-cycle N is enriched at the cost of C and O. One 
would therefore expect a CN-CH anticorrelation if CNO-processed material is 
dredged up to the stellar surface. The so-called first dredge-up, however,
is not able to explain the observed abundance patterns of light elements 
in RGB stars, especially for metal-poor stars that do not
possess deep enough convective envelopes according to standard models \citep[see references in][]{G/S/C:04}. An additional mixing episode
is needed to explain those patterns. This can either be rotation-induced 
mixing \citep[e.g.,][]{S/M:79, C:95} or 
so-called `canonical extra-mixing' \citep{D/VdB:03}.
These mechanisms naturally explain the [C/Fe]-[N/Fe] anticorrelation 
observed in RGB stars, however will not work for stars below the RGB bump due
the increased molecular weight barrier \citep[e.g.,][]{I:68}. 
Based on low resolution spectroscopy, various studies showed that the CN-band 
strength is a good indicator for the
[N/Fe] abundances whereas CH traces [C/Fe] \citep[e.g.,][]{S/S:96}. 
As a consequence, the CN bimodality and the CN-CH anticorrelation 
observed on the upper RGB stars of many clusters are often interpreted as a 
result of deep mixing that takes place in certain stars while not in other 
stars.  

2) The `primordial' and `self-enrichment' scenarios:
In both cases the abundance variations are not due to internal stellar evolutionary effects. 
The 'primordial' scenario assumes that there exists a `primordial floor of abundance
variations' \citep{G/S/C:04} that was in place when the star cluster formed
(i.e., an inhomogeneously mixed molecular cloud).
In the 'self-enrichment' scenario the abundance variations are caused by successive
generations of stars that formed within the same star cluster.
Theoretical nucleosynthesis models show that the observed abundance mix can be provided
either by intermediate-mass (4--5 M\sun) asymptotic giant branch (AGB) 
stars \citep[e.g.,][]{C/DC:81, V/DA:01, D/H:03}, or by fast rotating
massive (20--120 M\sun) stars \citep[e.g.,][]{M/M:06, D/M:07}. 
Both types of objects expel their 
ejecta via slow stellar winds, which is important in order to not sweep out 
the gas from which the second generation shall be formed.
There are mainly two ways how the enriched stars got to their peculiar 
abundance pattern: either, the AGB ejecta mixed well with the intracluster
medium out of which the second generation formed within the cluster \citep{C/DC:81}.
Or, the AGB ejecta polluted the surfaces of a certain 
fraction of already existing stars with well-developed radiative cores \citep[e.g.,][]{DA/G:83, T/J:02}.
The pollution scenario, however, has difficulties to explain the sharp
bimodality of CN abundances and the similarity of abundance patterns of
evolved as well as unevolved stars.

Lately, the evolutionary mixing scenario has been more and more challenged as 
correlations/anticorrelations among these elements and the range of variations 
of each element appear to be independent of stellar evolutionary states 
(with exception of enhanced depletion of C and O seen on the RGB) 
\citep[e.g.,][]{H/S/G:03}.
Recent spectroscopic studies near and below the main sequence turn-off (MSTO)
in the GCs M\,71, 47\,Tuc and NGC\,6752 showed that abundance variations are 
already present among stars that are expected to be unaffected from deep 
mixing mechanisms \citep[e.g.,][]{C:99.2, H/S/G:03, B/H:04}. 
This suggests that at least some of the abundance variations observed in 
evolved stars were present before the stars reached the RGB, i.e. mixing can 
not be the only driving mechanism of the observed abundance variations.

\begin{figure*}[t!]
\centering
\epsfig{figure=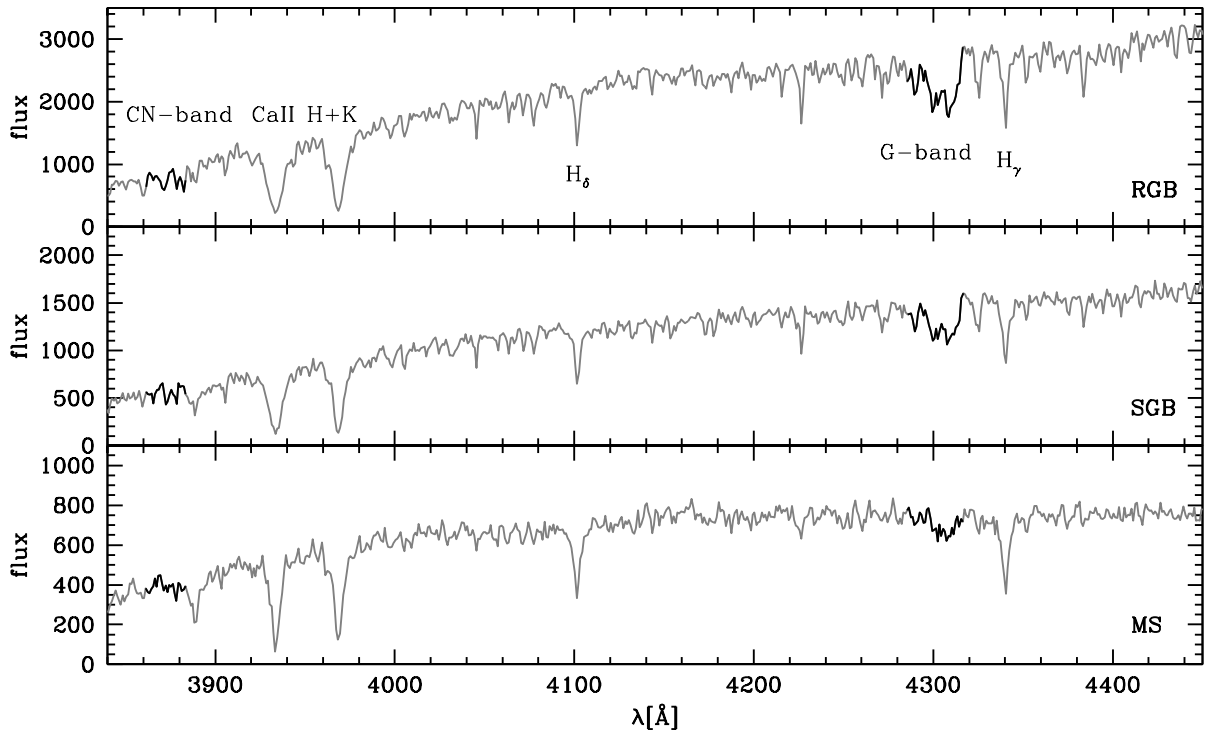,width=15cm,bbllx=25mm,bblly=130mm,bburx=200mm,bbury=246mm}
\caption{\label{spec} Typical spectrum of a RGB (top panel), a SGB (middle panel), and a
MS (bottom panel)
star in the globular cluster NGC\,288. The regions of the measured indices are marked by
darker lines. Furthermore the positions of the prominent CaII H and K and Hydrogen lines
are indicated in the top panel.}
\end{figure*}

The 'self-enrichment' scenario also is strengthened by the recent findings of
multiple subgiant branches (SGB) and main sequences (MS) in several massive 
GCs \citep{B/P:04, P/B:07}, which require stellar populations with distinct 
abundance patterns (and ages) within the clusters. Interestingly, the multiple
SGBs and MSs can best be explained by a large helium enhancement in the 
second/third subpopulation of a cluster \citep{DA/B:05}, which is consistent 
with the expected abundances of ejecta from intermediate-mass AGB stars 
\citep[e.g.,][]{DA/C:02}. Actually, these AGB stars need not have been members
of the same star cluster. \citet{B/C:07} recently suggested that
massive GCs might have formed in low mass dwarfs embedded in a dark matter 
halo. In this scenario, the second/third generation of stars then was created
out of ejecta from the external `field' AGB stars.
Since products of H burning are realeased by fast rotating massive stars in slow winds, also this
class of stars may provide the He-enhancement required to explain multiple
sequences observed in globular clusters.

Coming back to the overall CNO abundances, the work by \citet{S/S:96} has 
shown that the total 
$[({\rm C+N+O)/Fe}]$ for giants in the globular clusters M\,3 and M\,13 is 
the same for both CN-strong and CN-weak stars, which would be expected from 
deep mixing, dredging up CNO processed material to the stellar surfaces.
Thus although mixing effects are not existent in unevolved stars they 
seem to play a role for red giants when studying the CN and CH bands.
The challenge is to disentangle the primordial contribution to the
C,N abundances from the one resulting from normal evolutionary changes. 
On the one hand, some basic evolution of low mass population\,II stars is clearly
a common feature in both field and cluster stars.
\citet{S/M:03} showed that halo field giants and
globular cluster giants share the same pattern of declining C as a function of
increasing magnitude. The same two mixing mechanisms (first dredge-up and a
second mixing episode after the RGB-bump) are acting in all population\,II giants.
On the
other hand, field stars behave very differently from cluster stars as far as
"heavier" light elements (namely O, Na) are concerned \citep{G/S:00}.

If the environment in which a cluster formed (e.g., in the disk of a galaxy vs. the
center of a dark matter substructure) defines the enrichment history of a cluster,
the observed abundance patters would provide an indication of the origin of the cluster.
In his groundbreaking work \citeauthor{Z:85} proposed that the Galactic globular 
cluster system consists of various sub-systems \citep{Z:85, Z:93}: bulge/disk
(BD), old halo (OH), young halo (YH) globular clusters. He furthermore 
suggested that most YH clusters might have been accreted from satellite 
galaxies.
However, the Milky Way companions have been found to show, on average, 
systematically lower [$\alpha/{\rm Fe}$] ratios than Galactic halo stars and 
globular clusters \cite[e.g.,][]{S/C:01, F:02, P/V:05, S/B:07}.
Hence the present-day dwarfs do not seem to have contributed in a significant
way to the build-up of the Galactic halo and to the YH clusters.

The aim of this work is to gain further insight into the mechanism responsible
for the strong CN enhancement in some stars. We therefore concentrate on 
regions in the color magnitude diagrams (CMDs) where stars are believed to be
unaffected by mixing effects, i.e. stars on the MS, MSTO, SGB, and lower RGB.
In particular, we investigate
whether there is a dependence of the CN enhancement on the overall globular cluster
properties and/or the sub-class they are belonging to.
We investigate if CN-CH variations are different in genuine halo clusters as 
compared to possibly accreted globular clusters.

This article is structured as follows.
Section~\ref{MWGC:data} describes our data and their reduction.
Section~\ref{MWGC:CNCH} explains the measurements of the CN and CH band 
strength and the definition of the cyanogen excess parameter. 
Sections~\ref{MWGC:anti} and \ref{MWGC:trends} present the investigation of 
the CN/CH anticorrelation and the search for correlations between other cluster properties and the number ratio of CN-strong/CN-weak stars. The final section 
\ref{MWGC:sum} gives our summary and conclusions.

\section{Observations and data reduction}\label{MWGC:data}
\begin{figure*}[t!]
\centering
\begin{tabular}[]{c c c}
\vspace*{1.0cm}\\
\hspace*{-0.6cm}\psfig{figure=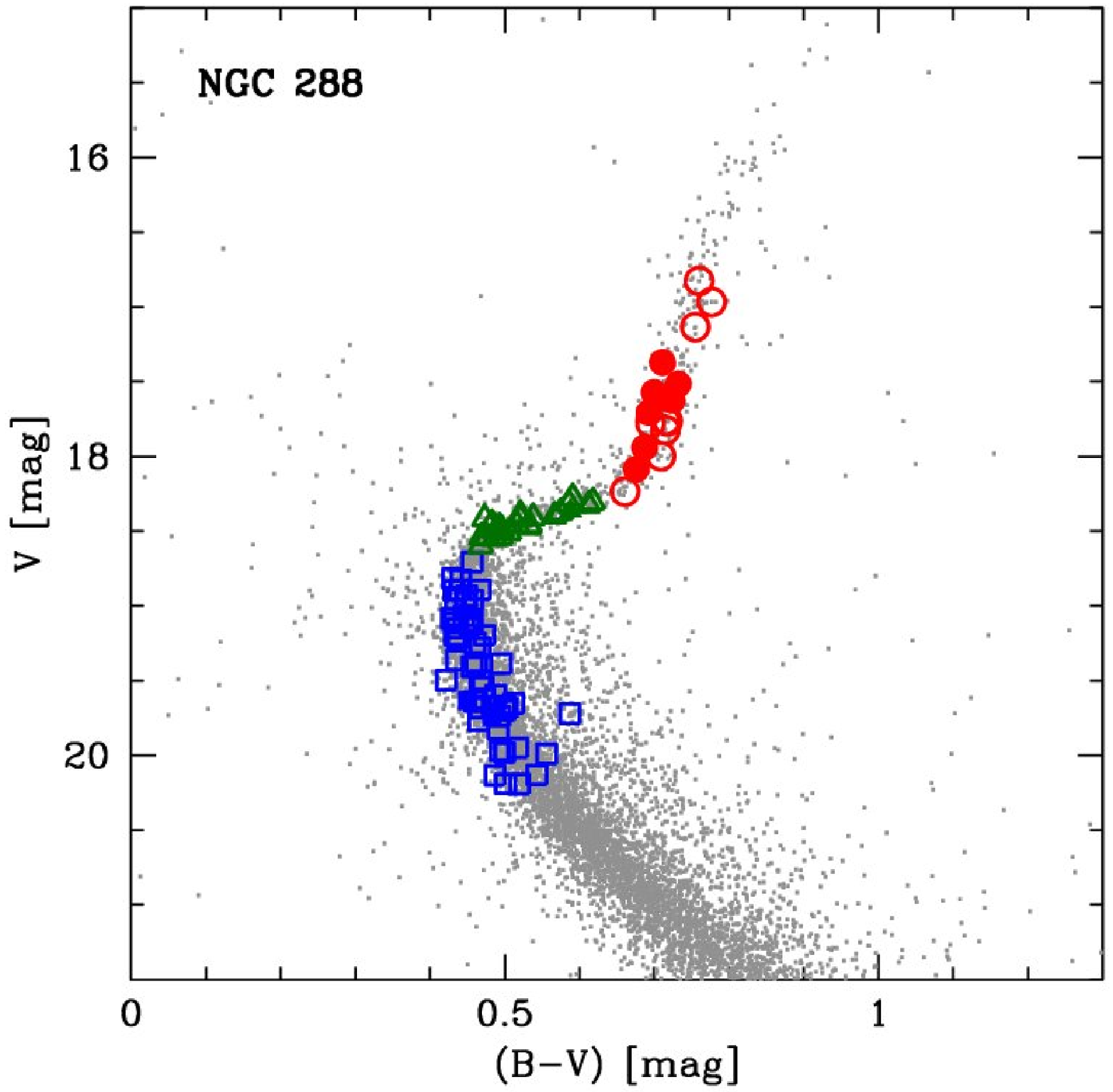,height=5.6cm,bbllx=9mm,bblly=20mm,bburx=195mm,bbury=200mm}&
\hspace*{-0cm}\psfig{figure=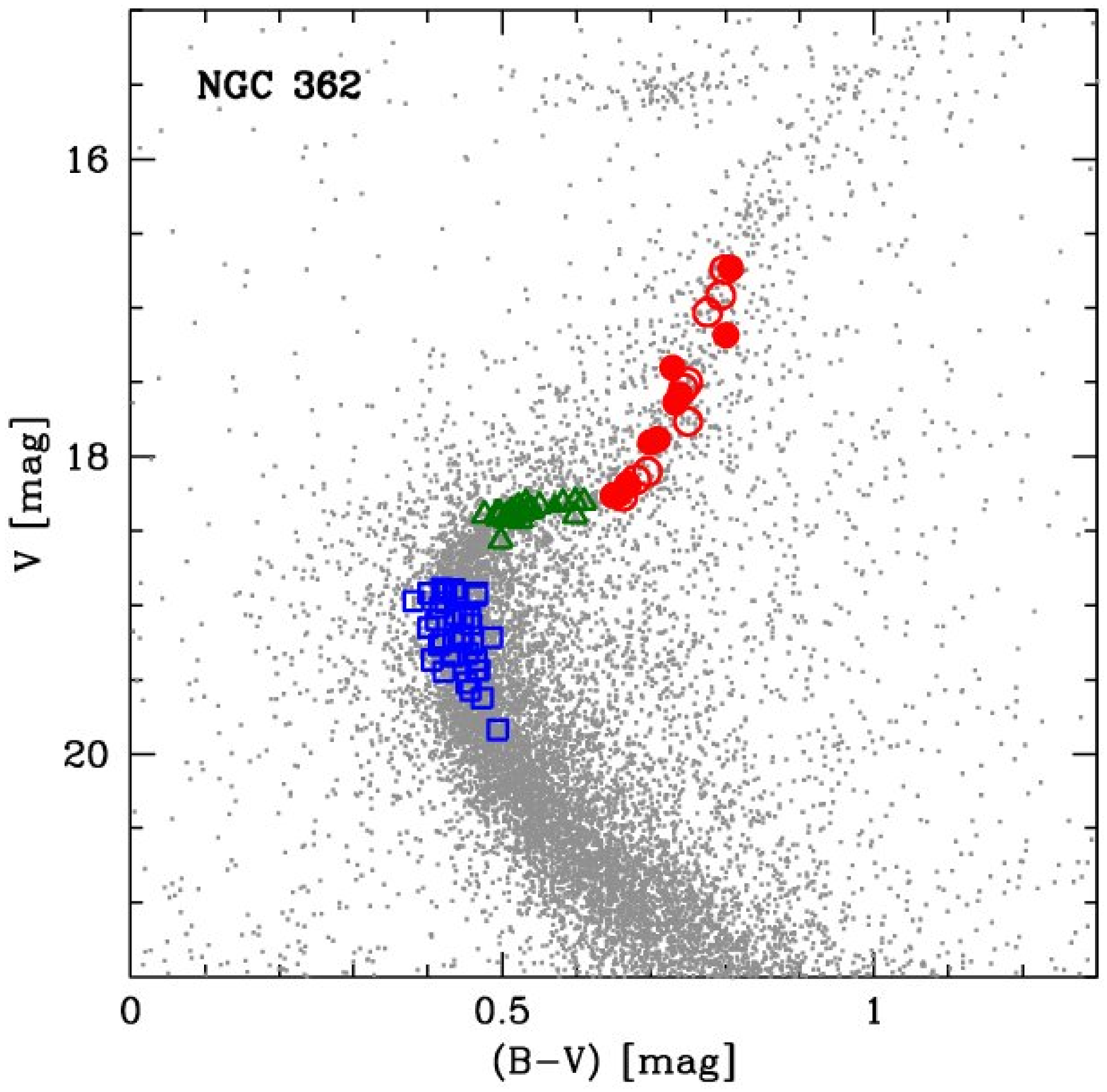,height=5.6cm,bbllx=9mm,bblly=20mm,bburx=195mm,bbury=200mm}&
\hspace*{-0cm}\psfig{figure=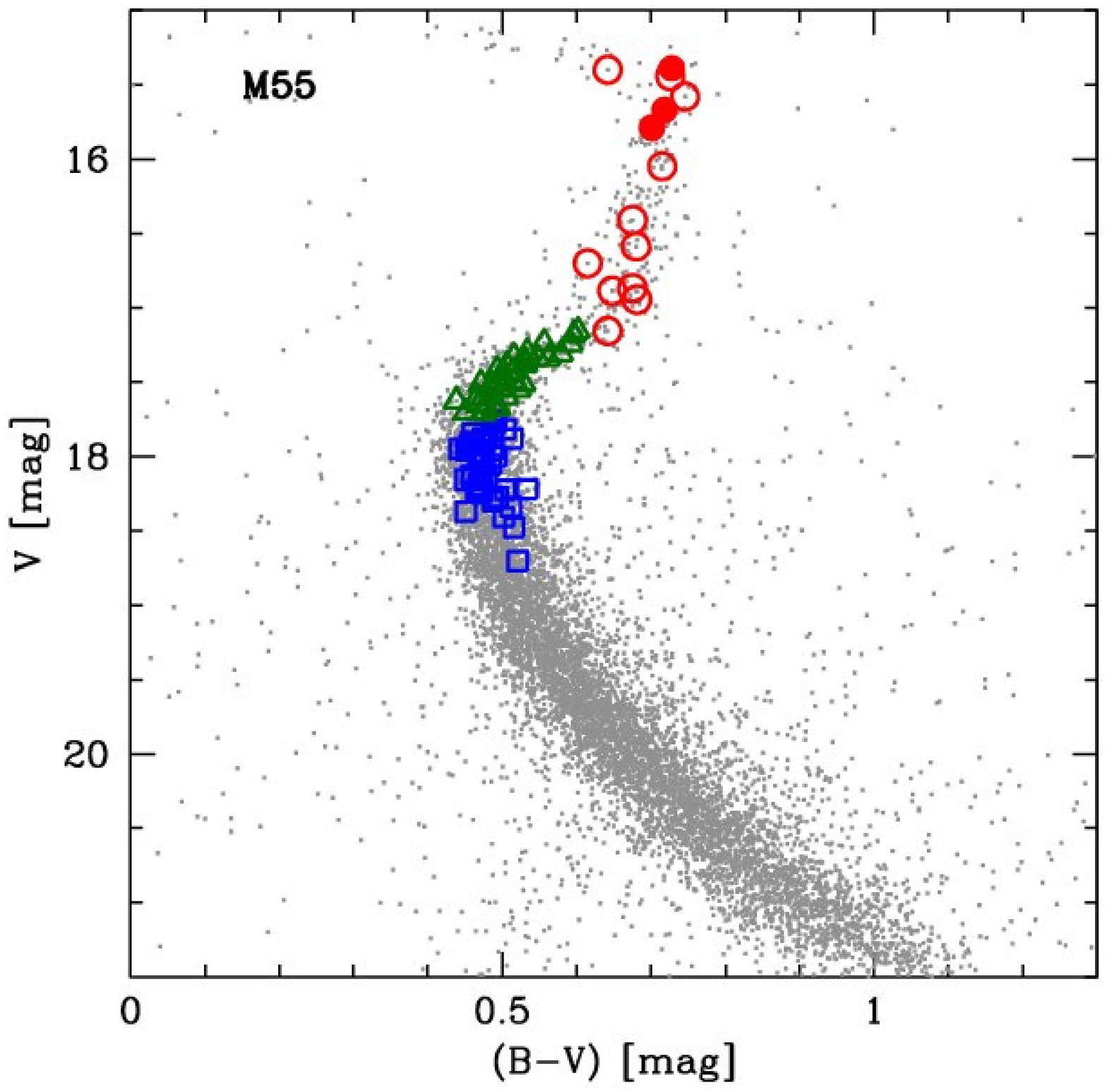,height=5.6cm,bbllx=9mm,bblly=20mm,bburx=195mm,bbury=200mm}
\vspace*{1.0cm}\\
\hspace*{-0.6cm}\psfig{figure=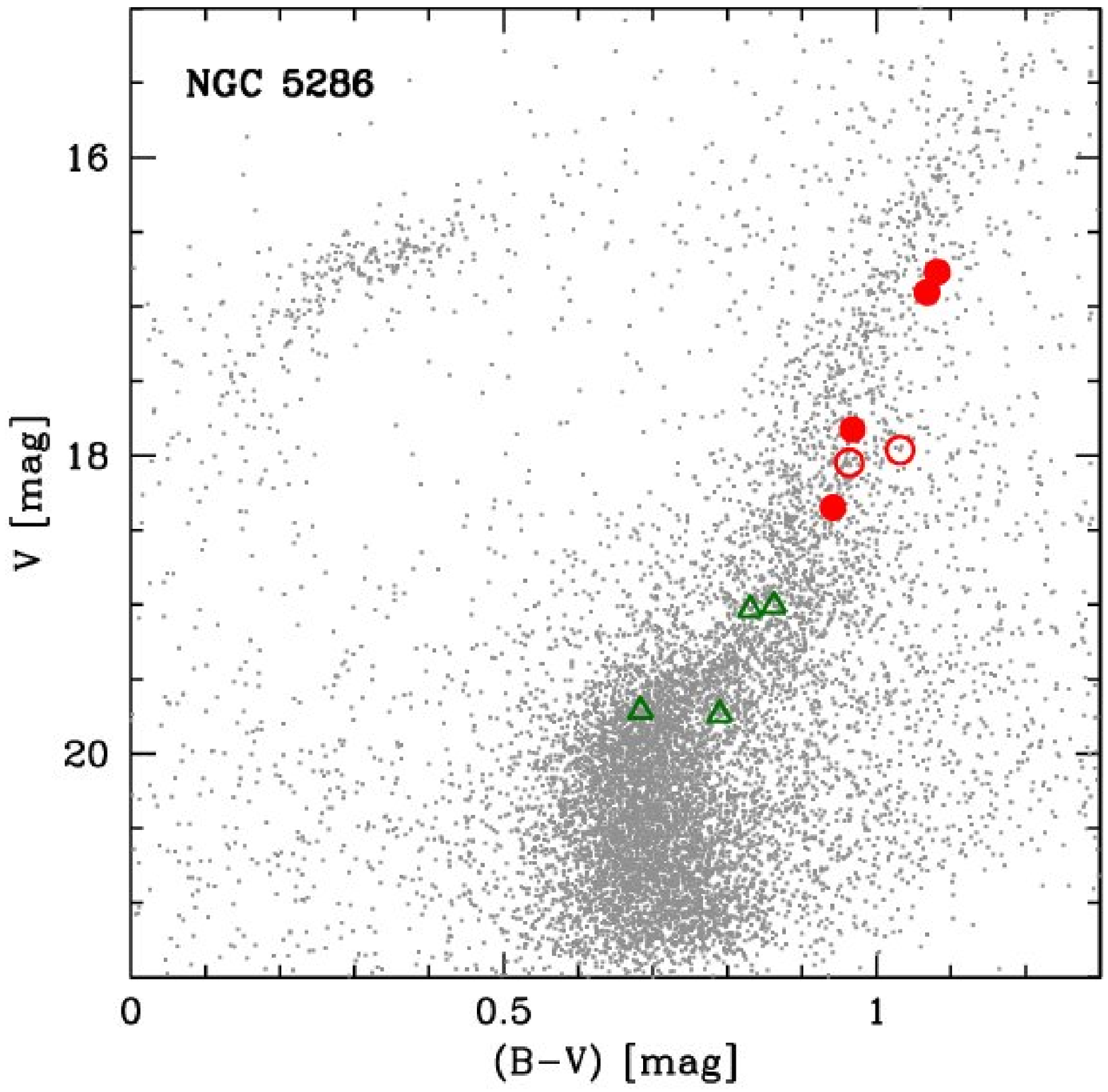,height=5.6cm,bbllx=9mm,bblly=20mm,bburx=195mm,bbury=200mm}&
\hspace*{-0cm}\psfig{figure=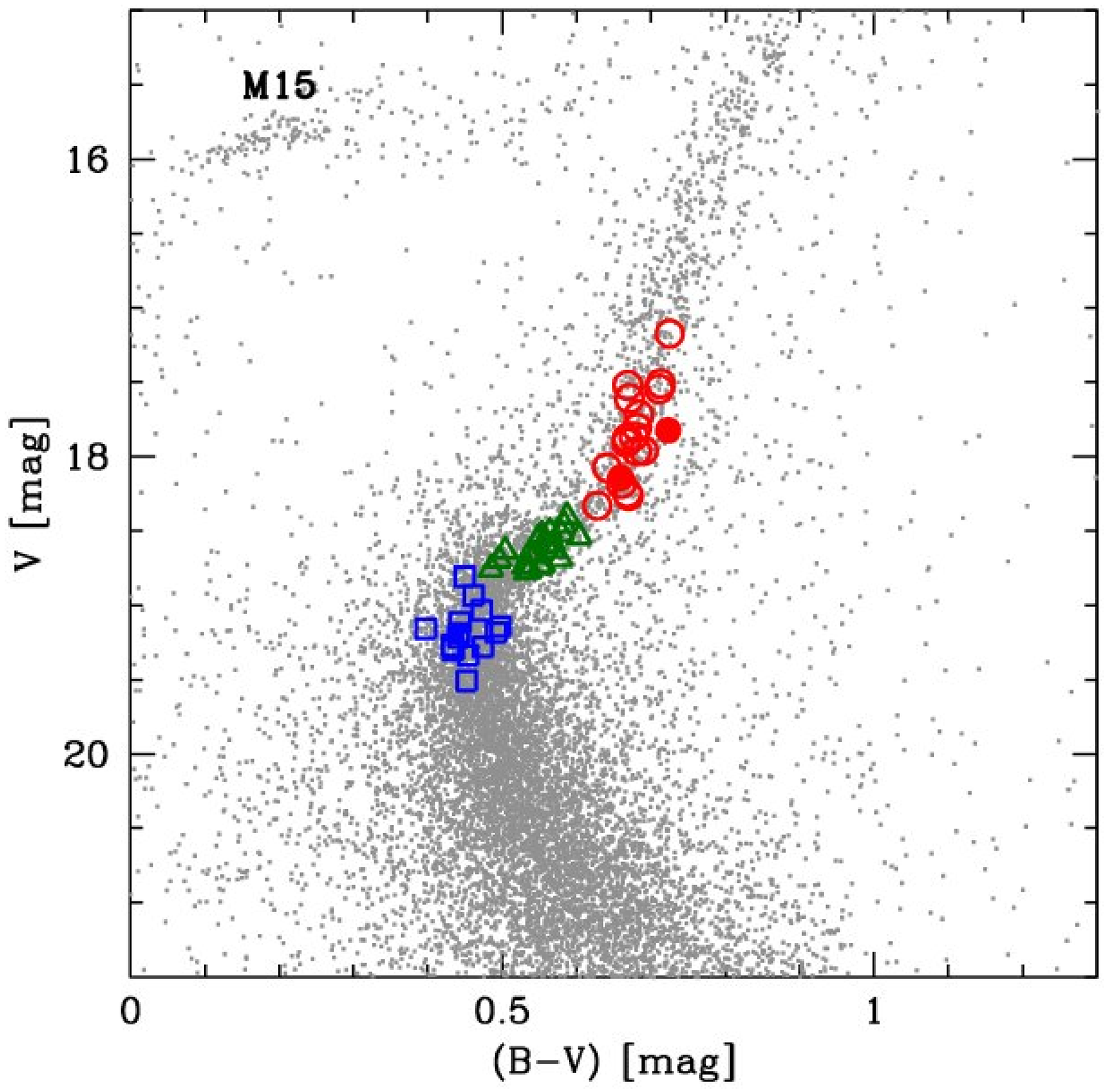,height=5.6cm,bbllx=9mm,bblly=20mm,bburx=195mm,bbury=200mm}&
\hspace*{-0cm}\psfig{figure=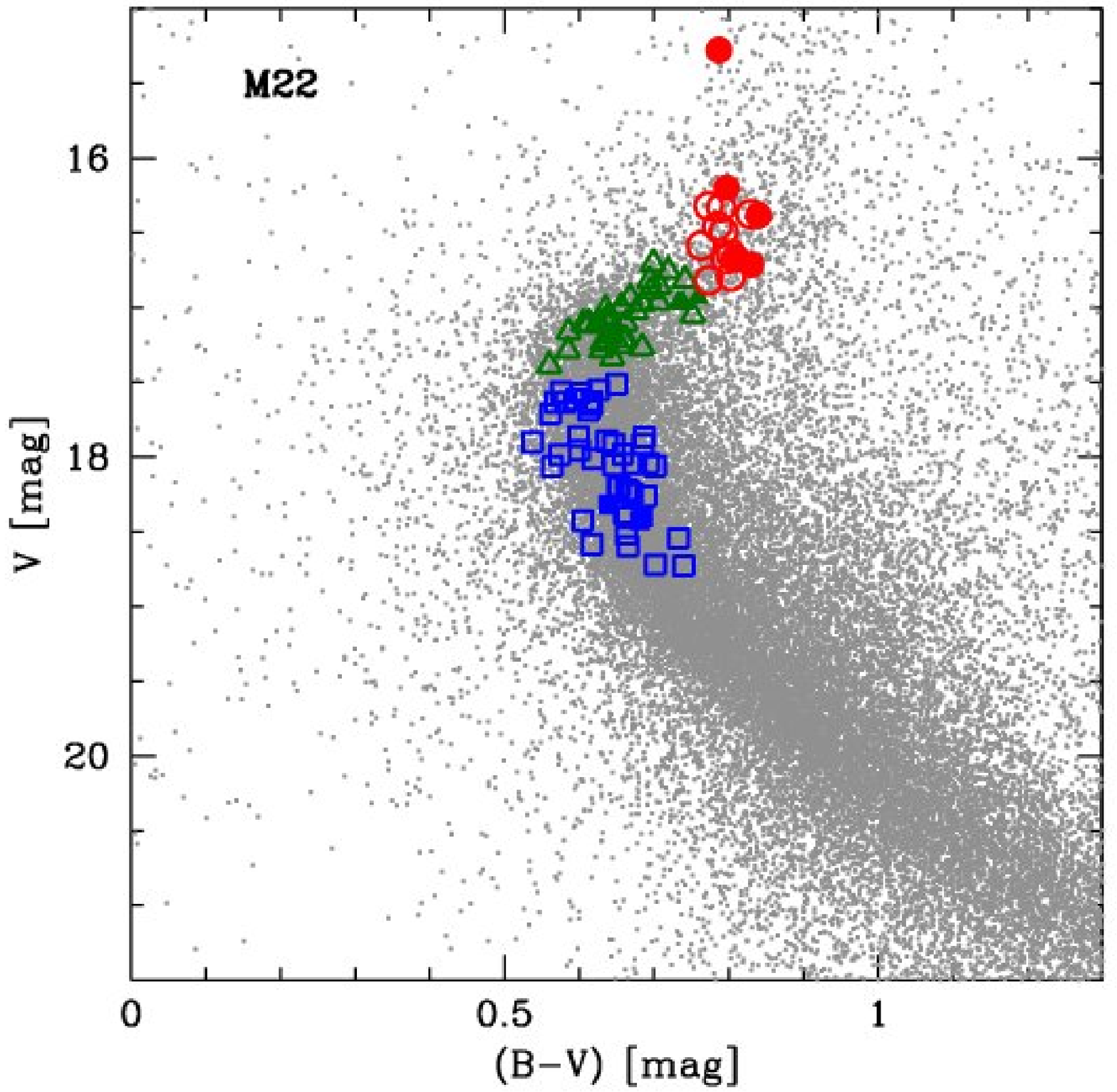,height=5.6cm,bbllx=9mm,bblly=20mm,bburx=195mm,bbury=200mm}
\vspace*{1.0cm}\\
\end{tabular}
\begin{tabular}[]{c c}
\hspace*{-0cm}\psfig{figure=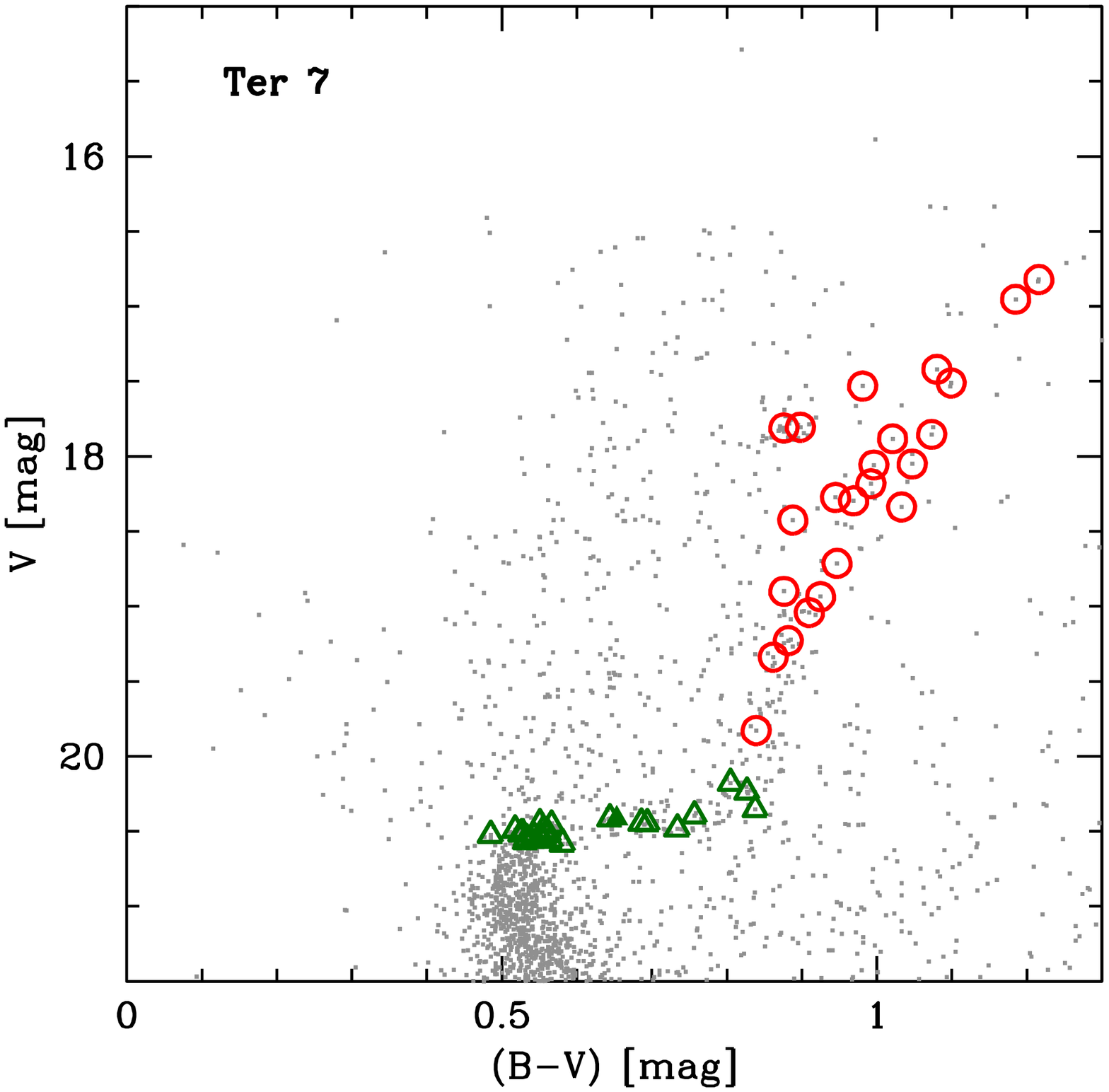,height=5.6cm,bbllx=9mm,bblly=60mm,bburx=195mm,bbury=246mm}&
\hspace*{-0cm}\psfig{figure=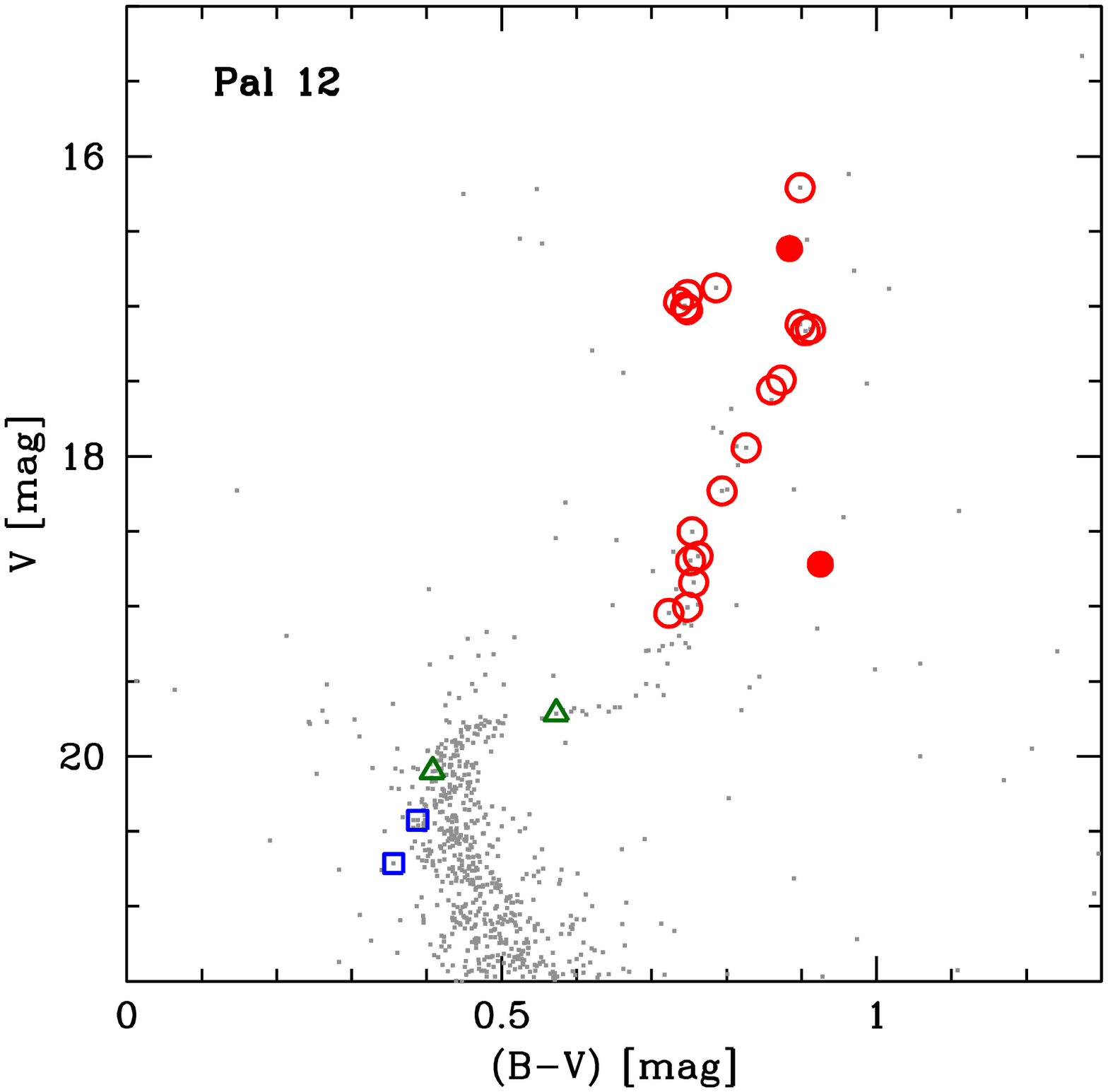,height=5.6cm,bbllx=9mm,bblly=60mm,bburx=195mm,bbury=246mm}
\vspace*{0.9cm}\\
\end{tabular}
\caption{\label{cmd} The color magnitude diagrams for the globular clusters 
in our sample. Those stars for which line strength measurements are available 
are marked in color. We distinguish between stars of different evolutionary 
states. MS stars are indicated by blue squares, SGB stars by green triangles, 
and RGB stars by red circles. CN-weak and CN-strong stars are denoted by open 
and filled symbols, respectively. Note that in all diagrams the calibrated 
pre-image $B$ and $V$ magnitudes are shown. Our sample comprises clusters 
spanning a large range in metallicity ($-2.26<$[Fe/H]$<-0.58\unit{dex}$). The 
clusters Palomar\,12 and Terzan\,7 are believed to be part of the Sgr dSph, 
which is currently being disrupted by its tidal interaction with the Milky 
Way.}
\end{figure*}

The spectroscopic data were obtained in May 2002 and July 2004 at the VLT/UT4 at ESO/Paranal (Chile) with the multi-slit spectroscopy instrument FORS2/MXU.
FORS2 provides a field of view of $6\farcm 8\times 6\farcm 8$.
The observations of M\,55 were obtained in 2002 and were also used for calibration purposes in a study of \wcen\ \citep{H/K:04, W/H:05, K/H:06}.
The observations obtained in 2004 were dedicated to CN and CH
measurements in seven further Galactic globular clusters (M\,15, M\,22, M\,55, NGC\,288, NGC\,362, NGC\,5286, Palomar\,12, and Terzan\,7) spanning a large range in metallicity ($-2.26<$[Fe/H]$< -0.58 \unit{dex}$).
Two of the clusters (Palomar\,12 and Terzan\,7) are suggested to have originated from the Sagittarius dwarf spheroidal (Sgr dSph) galaxy \citep{B/F/I:03, S/B:05}.

For both observing runs, the candidate stars for the spectroscopy were selected from pre-images in Johnson $B$ and $V$.
We selected target stars from the upper MS, the SGB, and the lower RGB in the cluster CMDs.
On the RGB we focused on stars fainter than the RGB bump, the point where deep mixing is believed to set in \citep{S/M:79, C:95}.

We chose the grating with the ESO denotation 660I+25 (second order) with a dispersion of $0.58 \unit{{\AA} \; pix^{-1}}$.
The spectral region covers $\sim 3700$ to $5800\unit{\AA}$ including the CN band at $3885\unit{\AA}$ and the G-band at $4300\unit{\AA}$.
The final actual wavelength coverage depends on the location of the star/slit on the mask with respect to the dispersion direction.
Typically we defined one slit mask per region of the CMD per cluster, containing $\sim$50--70 slits.
We selected slit lengths of 4--8 $\arcsec$ to make local sky subtraction possible.
The slit width was fixed to $1\farcs0$.
The total exposure time per mask varied between 360 and $5400\unit{s}$ depending on the cluster and the brightness of the target stars.
To facilitate cosmic ray removal the observations were split into multiple (2--3) exposures.
The central coordinates of the observed fields as well as the total exposure times are listed in Table~\ref{tab1}.
In addition to the science exposures, we obtained bias, flatfield and wavelength calibration observations.

\begin{table}[b!]
\caption{\label{tab4} Reddening, distance modulus and photometric parameters
of the MSTO for our sample GCs}
\begin{center}
\begin{tabular}[l]{ccccc}
\hline\hline
 Cluster & $E_{B-V}^a$ & $(m-M)_V^a$ & $ V_{MSTO}$ & $(B-V)_{MSTO}$ \\
 \hline
 NGC\,288       & 0.03	& 14.83 & 18.90$^b$ & 0.46$^b$	 \\	
 NGC\,362	& 0.05	& 14.81 & \multicolumn{2}{c}{similar to NGC\,288$^c$} \\
 NGC\,5286	& 0.24	& 15.95 & 20.05$^d$ & 0.73$^d$	 \\
 M\,22		& 0.34	& 13.60 & 17.70$^e$ & 0.75$^e$  \\
 Ter\,7		& 0.07	& 17.05 & 20.96$^f$ & 0.52$^f$  \\
 M\,55		& 0.08	& 13.87 & 17.89$^g$ & 0.50$^g$ \\
 M\,15		& 0.10	& 15.37 & 0.50$^h$ & 19.40$^h$	 \\
 Pal\,12	& 0.02	& 16.47 & $\sim$20.5$^i$ & 0.452$^i$ \\
\hline\hline
\end{tabular}
\end{center}
\vspace{-0.2cm}
\footnotesize{$^a$\citet{H:96}
	      $^b$\citet{A/L:97}
	      $^c$\citet{B/P:01}
	      $^d$\citet{S/I:95}
	      $^e$\citet{S/K:95}
	      $^f$\citet{B/C:95}
	      $^g$\citet{A/L:92}
	      $^h$\citet{D/H:93}
	      $^i$\citet{S/H:89}\\}
\end{table}

The photometric data are based on the pre-image observations of the target 
fields in the $B$ and $V$ band, taken several months prior to the spectroscopic
observations with FORS2 at the VLT/UT4.
The identification and psf-photometry was performed on the
pipeline reduced images (provided by ESO) using the the \texttt{IRAF} package
\texttt{DAOPHOT}. $B$ and $V$ magnitudes were matched to create the CMDs.
For this work, a precise photometric calibration is not necessary since we are
mainly interested in a comparative study of stars in different evolutionary 
states, which can easily be identified in the CMDs. A rough calibration was done
by adjusting the zeropoints such that the MSTO $(B-V)$ colors and $V$ 
magnitudes taken from the literature were matched (see Table~\ref{tab4}).

Based on the location in the CMDs we assigned stars to the MS, SGB, and RGB.
Figure~\ref{cmd} shows the CMDs for all clusters in our sample. The stars with
available spectra are symbol-coded according to their position in the CMD. 
Only those stars are shown that were identified as radial velocity members and
that passed our quality check of the spectra. For the two Sgr clusters Ter\,7 
and Pal\,12 some stars near the RGB bump have been observed. These stars are 
included in the Figures~\ref{cmd}, \ref{CNMV}, and \ref{CNCH} but neglected in
the further analysis.

The data reduction was carried out using standard routines within 
\texttt{IRAF}\footnote{IRAF is distributed by the National Optical Astronomy 
Observatories, which are operated by the Association of Universities for 
Research in Astronomy, Inc., under cooperative agreement with the National 
Science Foundation.}. This included bias correction and flatfielding. The 
cleaning for cosmic rays was done with \texttt{bclean} from the 
\texttt{STARLINK} package. Before sky subtraction was performed the spectra 
from the individual exposures were stacked to improve the signal-to-noise 
ratio. In most cases, object and sky could be extracted from the same slit.
The wavelength calibration was achieved using the emission spectra of the 
He-Ne-Hg-Cd arc lamps taken after each set of observations. Note that the 
final spectra were neither flux-calibrated nor normalized by the continuum.
All spectra were binned to a spectral scale of $1\unit{\AA pix^{-1}}$.
Considering the seeing the final spectral resolution (FWHM) for narrow 
lines is $\sim 2\unit{\AA}$.
Typical spectra of a RGB, a SGB and a MS stars in NGC\,288 are shown in
Fig.~\ref{spec}.

For all spectra we measured radial velocities by cross-correlating them with 
five high quality template spectra taken from the \wcen\ dataset using 
\texttt{IRAF/fxcor}. We adopted the mean value of the five measurements as the
radial velocity of the star and corrected for the measured Doppler shift. The 
scatter of the velocity measurements is of the order of $20\unit{km/s}$, which 
reflects the uncertainties given by the spectral resolution. In the resulting 
velocity distributions the globular clusters clearly stand out against the 
Galactic foreground. Possible non cluster member stars were identified by 
their radial velocities and rejected from the further analysis. In a final 
step, we examined each spectrum individually and rejected those spectra with 
bad quality (e.g., due to tracing errors). In total about 500 spectra are 
suitable for our analysis, whereof 120 spectra are from lower RGB stars.

Note that NGC 5286 and M22 have quite a high foreground extinction, and
probably differential reddening is broadening the giant branches (e.g. Richter
et al. 1999). Most of the radial velocity members of NGC 5286 lie on the red 
side of the RGB sequence which might reflect their biased location west of the
cluster centre (pointing of the spectroscopic mask).

In the Appendix, magnitude limited samples of cluster member stars that 
were used for our analyses are presented (Table~\ref{tabAPP}).
Only the brightest five stars of each cluster are contained in this table.
The full table of all cluster stars only is available in the online version 
of the article.

\section{CN and CH band strengths}\label{MWGC:CNCH}
For all spectra, we measured line indices covering the absorption features of 
the CN and CH molecules. For the CN and CH band strengths, we used the modified
S3839 and CH4300 indices as defined by \citet{H/S/G:03}:
\begin{equation}
{\rm S3839\, (\rm CN) = -2.5\;log \frac{F_{3861-3884}}{F_{3894-3910}}},
\end{equation}
\begin{equation}
{\rm CH4300= -2.5\;log \frac{F_{4285-4315}}{0.5F_{4240-4280}+0.5F_{4390-4460}}},
\end{equation}
where ${\rm F_{\lambda}}$ are the fluxes in the different bandpass regions.
Our error estimates assume Poisson statistics in the flux measurements.

\subsection{CN band strength}\label{sec:CN}
\begin{figure}[t!]
\centering
\epsfig{figure=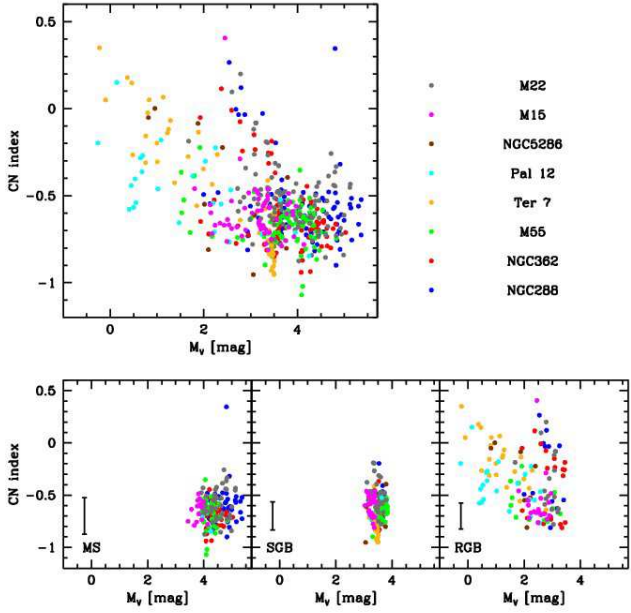,width=8cm,bbllx=25mm,bblly=0mm,bburx=200mm,bbury=200mm}
\caption{\label{CNMV} The distribution of the stars of the different clusters 
in the CN vs. $M_V$ diagram. The left upper panel illustrates the overall 
distribution of our sample stars. In the lower panel we distinguished between 
different evolutionary states of the stars. The color coding of the data 
points corresponds to stars from the different clusters as indicated in the 
figure legend in the upper right. Whereas for the MS all clusters show roughly
the same distribution, for the RGB the distribution shows a large scatter. For
the clusters NGC\,288 and NGC\,362 a bimodal distribution in CN band strength 
is visible. In the lower left corner of the bottom panels the median errors of
the measurements are shown.}
\end{figure}

In order to investigate the behavior of the strengths of the CN index as a 
function of evolutionary state (or stellar mass) we plotted CN against the 
absolute $V$ magnitude, $M_V$, for all clusters (Figure~\ref{CNMV}).
We adopted the distance moduli and extinction values of \citet{H:96}.

Looking at the whole sample of stars a wide spread in CN and a continuous 
increase of CN with decreasing $M_V$ can be seen in the upper panel of this 
figure. This is caused by the fact that the formation of molecules in stellar
atmospheres strongly depends on the effective temperature, \Teff , and the 
surface gravity,  \logg\, of the stars. The efficiency of CN formation is 
higher in stars with lower \Teff\ and lower \logg .
To further illustrate this effect we subdivided our sample into MS 
(\logg$\sim 4.5$, \Teff$\sim 6000\unit{K}$), SGB (\logg$\sim$4.5--3, 
\Teff$\sim$5000--6000$\unit{K}$), and RGB stars (\logg$\sim 3$, \Teff$\sim 
5000\unit{K}$). The different distributions for the different evolutionary 
states are shown in the lower panels. One can clearly see that the line 
strengths of CN on average increase as stars evolve from the MS to the RGB.
This can be understood by the augmented formation of molecules in cooler 
atmospheres.

Looking at the globular clusters individually one recognizes that they show 
very different behaviors in the \mv\ vs. CN diagram. Whereas for the MS and 
the SGB all clusters show roughly the same pattern, the distributions on the 
RGB deviate between the clusters. For some clusters like e.g., NGC\,288 and 
NGC\,362, we clearly see a bifurcation in CN band strengths as we reach the RGB.
Either part of the bifurcation contains roughly equal numbers of stars.
This is worth to keep in mind as the two clusters are a so called 
``second-parameter pair'': both clusters have similar metallicities but show 
a very distinct horizontal branch morphology. In NGC\,288, most of the core 
helium burning stars can be found on the blue horizontal branch whereas almost
no stars are located on the red part. Exactly the opposite is the case for 
NGC\,362. For this cluster the red part of the horizontal branch is densely 
populated. Some authors proposed that deep mixing and the consequently 
increased mass loss could be an explanation for the different horizontal 
branch morphologies as well as the observed abundance anomalies 
\citep[e.g.,][]{W/D:00}. For other second parameter pairs like e.g., M\,3 and 
M\,13, which also show differences in light abundance elements this might be a
possible explanation for the observed patterns.
Both clusters have similar ages and metallicities.
However, the RGB in M\,3 is dominated 
by CN-weak stars, whereas the majority of stars in M\,13 are found to be 
CN-strong \citep[e.g.,][]{S:81}.

Nevertheless, the fact that we do not observe significant differences in the 
CN distributions indicates that deep mixing cannot be a major cause of the 
horizontal branch morphology. Similarly, based on the CN and CH measurements
of stars in the second parameter globular cluster NGC\,7006, \citet{H/S/G:03b} argued 
against the hypothesis that CN-variations are directly correlated with the 
second parameter effect. They found the scatter in CN to be similar to those 
in other GCs of the same metallicity but different horizontal branch ratios.

\begin{figure}[t!]
\centering
\epsfig{figure=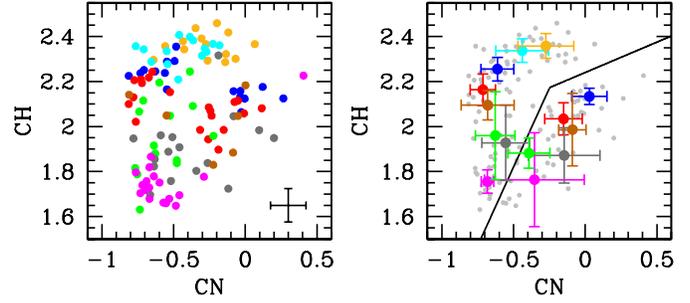,width=8cm,bbllx=25mm,bblly=150mm,bburx=200mm,bbury=246mm}
\caption{\label{CNCH} The distribution of the stars in the CN vs. CH diagram 
for the RGB star in our sample clusters. The stars of different clusters are 
indicated by different colors as listed in Fig~\ref{CNMV}. The left panel 
illustrates the overall distribution of our sample stars, with a typical error
given in the lower right corner. The solid line indicated a possible 
differentiation between CN-strong and CN-weak stars in this diagram, drawn by
eye. In right panel we calculated the mean CN and CH of both, the CN-strong 
and CN-weak stars. The original overall distribution is plotted in grey, while
the mean values are color-coded as defined before.}
\end{figure}

In contrast to NGC\,288 and NGC\,362, the clusters Ter\,7, Pal\,12, and M\,55
seem to exhibit no or only very few stars with strong CN band strengths. In 
the clusters NGC\,5286, M\,22, and  M\,15, stars can be found on both the 
CN-weak and the CN-strong regime in this diagram. For M\,15 and M\,22, the 
majority of the stars are associated with the CN-weak group. For NGC\,5286, 
we have only six measurements. Four of these stars are found to be CN-strong 
and two CN-weak.

We cannot assess whether similar abundance variations on the SGB and the MS 
region are not present or can not be detected due to a too weak signal caused 
by the higher effective temperatures of these stars. The observed scatter in 
the CN measurements of MS and SGB stars (rms $\sim$ 0.13 and 0.14, 
respectively) is found to be of the same order as the errors in the index 
measurements (0.17 and 0.13, respectively).

CN as a double-metal molecule is easier to observe in more metal-rich 
clusters due to the stronger equivalent widths at higher metallicities.
Our work as well as former studies on the RGB show that whatever 
process is responsible for the formation of the CN-strong stars, it seems to 
occur in the majority of Galactic globular clusters. In contrast to this, in 
both fairly metal-rich Sgr dSph clusters (Pal\,12 and Ter\,7) we found no sign
for this process to be present. All stars in these clusters are located in the
CN-weak branch in Fig.~\ref{CNMV}. From the fact that, if present, CN-strong 
stars should show up easily in these clusters we infer that they actually 
lack those stars. This suggests that probably the environment in which the 
clusters formed had an effect on the presence or absence of the CN variations. 
However we point out that possible effects of the small sample size cannot be
ruled out.
Preliminary results from low resolution spectra of seven giants in Arp\,2 and Ter\,7
suggest star ot star variations in the CN band strength in these clusters \citep{B/M/S:07}.

\subsection{CN vs. CH}\label{sec:CNvsCH}
The CN vs. CH diagram also allows us to study CN bimodalities. In 
Fig.~\ref{CNCH} (left panel) we plot the measured CN vs. the CH band strengths
for the RGB stars in our sample clusters. The overall patterns found for RGB 
stars in Fig.~\ref{CNMV} also show up in Fig.~\ref{CNCH}.
A clear bifurcation into two branches is detected in the CN vs. CH diagram for 
stars on the RGB. NGC\,288 and NGC\,362 show the strong bimodality in the 
distribution of CN line strength, seen before. In contrast, the RGB data 
points of Ter\,7 and Pal\,12 again are both located on the CN-weak branch in 
Fig~\ref{CNCH}. Due to their fairly high metallicities these clusters are 
found in the CH-strong regime in this diagram. It seems as if the two Sgr 
clusters are more homogeneous in their CN abundances than the Galactic globular 
clusters in our sample.

Interestingly, the stars in M\,15, which showed no indication for a bifurcation
in Fig.~\ref{CNMV}, show a weak indication
of a bimodal distribution (two clumps separated at CN$\sim-0.6$) in the CN-CH plane (Fig.~\ref{CNCH}). However, 
this needs further confirmation since the observational errors of such weak 
lines are large compared to the separation of the two clumps. Assuming that 
the clump at CN$=-0.5$ and CH$=1.65$ defines the CN-rich population, 
this would strongly change the number ratio of CN-strong to CN-weak stars in 
M\,15 (see next section). In order to further illustrate the dichotomy in this
plot we separated CN-strong from CN-weak stars (see Fig.~\ref{CNCH} right 
panel) and calculated the mean CN- and CH-indices for each sub-population in 
the different clusters (large dots).

As we introduced earlier, one of the scenarios proposed to explain the 
variations in C and N in RGB stars in globular clusters is the dredge-up of 
material processed in the CNO cycle. In our case, the origin of the observed 
patterns/bimodalities can not only lie in such mixing effects as the analyzed 
stars are considerably fainter than the red bump at which the deep mixing 
mechanism is expected to set in. Although we did not find evidence for CN 
bimodalities among our SGB and MS stars 
\citep[cf.][]{H/S/G:03} we favor a scenario in which the cluster formed out of
chemically inhomogeneous material that was polluted by the outflows of fast rotating
massive stars or AGB stars \citep[e.g.,][]{C/DC:81, V/DA:01, D/M:07}.

\subsection{Cyanogen excess parameter ($\delta$CN)}\label{sec:dCN}
\begin{figure}[t!]
\centering
\epsfig{figure=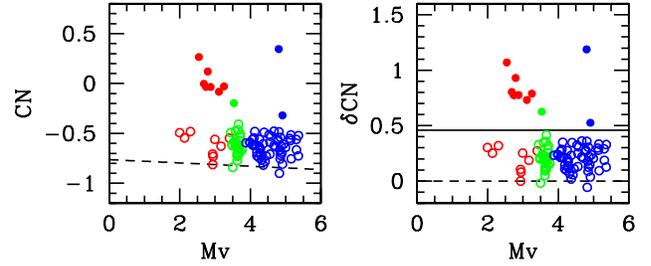,width=8cm,bbllx=25mm,bblly=150mm,bburx=200mm,bbury=246mm}
\caption{\label{bimo} Left: The \mv\ vs. CN diagram for the RGB (red), SGB 
(green) and MS (blue) stars for the cluster NGC\,288. The bimodal distribution
is clearly visible. The dashed line illustrates the lower envelope fitted to 
this distribution. Right: Plotted is \mv\ vs. the CN excess parameter \dcn . 
Stars with \dcn$>$0.46 are defined as CN-strong and indicated by filled 
circles. CN-weak stars are indicated by open circles. The solid line indicates
the separation between CN-strong and CN-weak stars.}
\end{figure}

As a measure to quantify the large range of CN line strengths we used a CN 
excess parameter (\dcn) similar to the one introduced by \cite{N/S:81}.
This minimizes the effects of effective temperature and surface gravity 
existent in the CN measurements. The \dcn\ parameter is defined as the CN 
strength with respect to a baseline. This baseline is defined by the lower 
envelope fitted for each individual cluster in the CN vs. $M_V$ distribution.
The left panel in Fig.~\ref{bimo} illustrates the baseline fit and right panel
shows the resulting \dcn\ vs. $M_V$ distribution for the cluster NGC\,288. 

\begin{figure}[]
\centering
\scalebox{1.28}[1.28]{\includegraphics*[viewport=30 45 499 547]{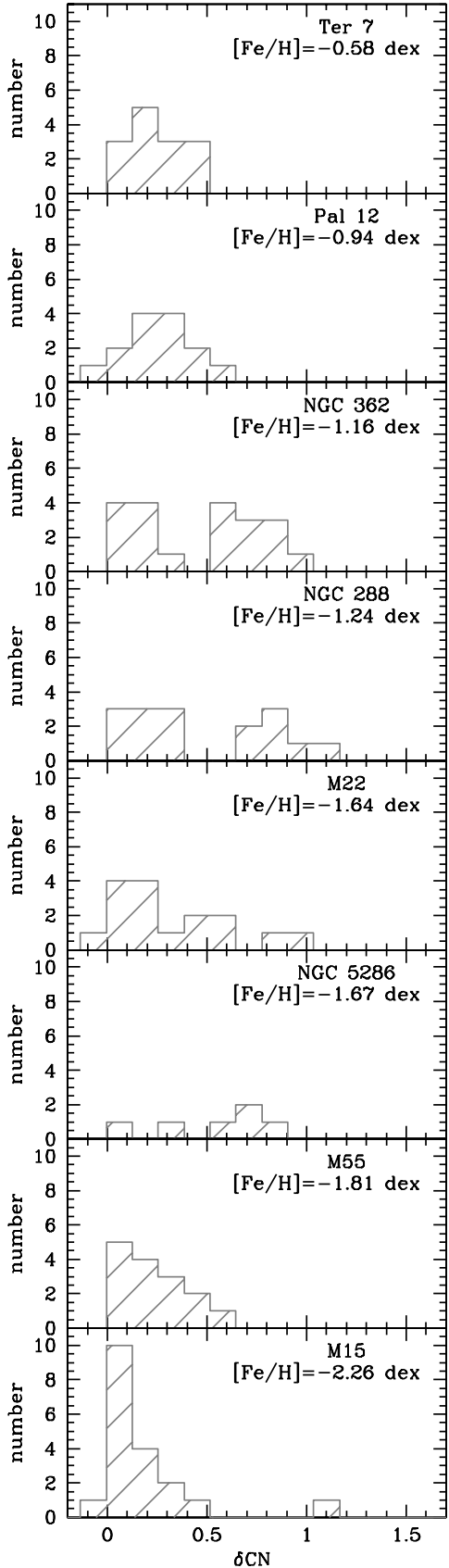}}
\caption{\label{histo_all} Distributions of CN-band strength in the RGB stars 
of our sample clusters. The histograms of the CN-excess parameter \dcn\ are 
plotted. The clusters are sorted by their metallicity. The two fairly 
metal-rich Sgr clusters are found in the two uppermost panels. The very CN-strong star in M15
lies slightly off the RGB and therefore is probably not a cluster member.}
\end{figure}

In the previous sections we saw that the bimodality is only clearly detected 
for stars on the lower RGB. Therefore, in the following we concentrate on this
part of the CMD. Fig.~\ref{histo_all} shows the histograms of the CN excess 
parameter for the RGB stars in all eight globular clusters in our sample, 
sorted by their metallicity. We selected a bin width of 0.13, comparable to 
the median uncertainties of the CN index for these stars.

Most of the metal-poor clusters (M\,15, M\,55, and M\,22) show a distinct main
CN-weak peak with a weak extension towards higher \dcn\ values. For NGC\,5286,
we observe a fairly flat distribution. However, due to the small sample size 
we cannot definitely comment on any distribution pattern. For NGC\,288 and 
NGC\,362, which have similar intermediate metallicities, the bimodal 
distribution clearly shows up in these plots. Both peaks are roughly equally 
pronounced. The two probable Sgr dSph clusters (Pal\,12 and Ter\,7) show a 
single peak and a fairly broad distribution around the CN-weak peak.
The spanned ranges in \dcn\ of about 0.45 and 0.53 for Ter\,7 and Pal\,12
are comparable with the ranges of 0.54 and 0.48 for M\,55 and M\,15 (excluding
the extremely CN-strong outlier in the last cluster) which at first glance
might point to a similar enrichment history of those clusters, despite their
very different metallicities and environments they live in. However, the
apparent broadness observed in M\,15 and M\,55 is mainly due to the metal-poor
nature of these clusters, resulting in larger errors in determining their CN
strength. In contrast, for the metal-rich clusters Pal\,12 and Ter\,7,
CN-strong stars and a bimodality are expected to clearly show up in these
diagrams, if present. This makes the chemical patters of the Sgr clusters
appear different from those of galactic clusters of similar metallicites
(e.g., 47\,Tuc) that show more pronounced CN spreads and/or bimodality.

\begin{figure}[t!]
\centering
\epsfig{figure=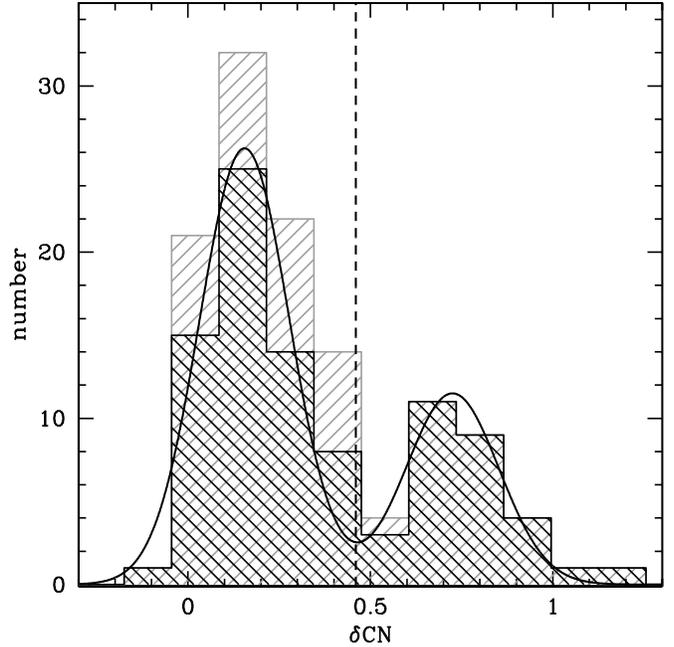,width=8cm,bbllx=25mm,bblly=60mm,bburx=200mm,bbury=246mm}
\caption{\label{histo_add} Combined histograms of the CN-excess parameter for 
the clusters in our sample. The grey histogram comprises all eight clusters. 
In the black histogram Pal\,12 and Ter\,7, which are believed to belong to the
Sgr dSph, are not included. Here we only consider stars on the lower RGB. 
\dcn\ shows a bimodal distribution, which was fitted by two Gaussians. The 
minimum between the two Gaussians was chosen as the criterion to differentiate
between CN-strong and CN-weak stars.}
\end{figure}

In Fig.~\ref{histo_add} we show combined histograms of the \dcn\ measurements 
of the clusters in our sample. We distinguish between a histogram of all eight 
clusters and one where we did not include the two Sgr dSph clusters, Ter\,7 and 
Pal\,12. In both cases a clear bimodal distribution is visible.
As Ter\,7 and Pal\,12 are of extragalactic origin 
we focused on the histogram based on six globular clusters. This 
distribution was used for the differentiation between CN-strong and CN-weak 
stars. We fitted a double Gaussian to the distribution and selected the 
minimum as the differential criteria between CN-strong and CN-weak stars.
CN-strong stars are then those that have a CN excess larger than \dcn$=0.46$.

In order to quantify the observed bimodality in the CN line strength, we 
determined the parameter r introduced by \cite{N:87}. It gives the number 
ratio of CN-strong to CN-weak stars.
\begin{equation}
{\rm r} = N_{\rm strong}/N_{\rm weak},
\end{equation}
Errors in r have been estimated from statistical uncertainties (adopting 
$\Delta N = \sqrt{N}$):
\begin{equation}
\Delta {\rm r} = r \sqrt{1/N_{\rm weak} + 1/N_{\rm strong}},
\end{equation}
where $N_{\rm weak}$ and $N_{\rm strong}$ give the number of CN-weak and 
CN-strong stars, respectively.

\begin{table}[]
\caption{\label{tab:ratio}CN number ratios}
\begin{center}
\begin{tabular}[l]{p{2.65cm} p{0.75cm} p{1.7cm} p{2.2cm}}
\hline\hline
Cluster & $\frac{\rm CN_{weak}}{\rm CN_{strong}}$ & r$_{\rm lower RGB}$ & r$_{\rm upper RGB}$ \\
\hline
NGC\,288          & 9/7  & $0.78\pm0.39$ &          \\	
NGC\,362          & 9/11 & $1.22\pm0.55$ & 2.46$^a$ \\
NGC\,5286         & 2/4  & $2.00\pm1.73$ &	    \\
NGC\,6656 (M\,22) & 10/6 & $0.60\pm0.31$ & 0.41$^b$ \\
Ter\,7	          & 14/0 & $0.00\pm0.00$ &          \\
NGC\,6809 (M\,55) & 13/2 & $0.15\pm0.12$ & 0.22$^c$ \\
NGC\,7078 (M\,15) & 18/1 & $0.06\pm0.06$ &          \\
                  & (12/7) & ($0.60\pm0.29$) &      \\
Pal\,12           & 13/1 & $0.07\pm0.07$ &          \\
\hline
NGC\,104 (47\,Tuc)&      & $1.90^d$      & 1.8$^d$  \\
NGC\,6839 (M\,71) & 13/9 & $0.69^e$      & 1.0$^e$, 0.63$^g$, 0.3$^h$\\
\hline
NGC\,1904 (M\,79) &      &               & 2.6$^b$ \\
NGC\,2808         &      &               & 2.4$^f$ \\
NGC\,3201         &      &               & 1.1$^f$ \\
NGC\,5272 (M\,3)  &      &               & 0.6$^f$ \\
NGC\,5904 (M\,5)  &      &               & 3.0$^f$ \\
NGC\,6121 (M\,4)  &      &               & 1.4$^f$ \\
NGC\,6171 (M\,107)&      &               & 1.4$^f$ \\
NGC\,6205 (M\,13) &      &               & 3.2$^f$ \\
NGC\,6254 (M\,10) &      &               & 0.5$^f$ \\
NGC\,6637         &      &               & 1.2$^f$ \\
NGC\,6752         &      &               & 1.6$^f$ \\
NGC\,6934         &      &               & 0.6$^f$ \\
NGC\,7006         &      &               & 2-3.5$^b$ \\
NGC\,7089 (M\,2)  &      &               & 3.8$^f$ \\
\hline \hline
\end{tabular}
\end{center}
\begin{center}
\vspace{-0.2cm}
\footnotesize{$^a$\citet{S/M:90}, $^b$\citet{H/S/G:03b},\\ $^c$\citet{N:87}, $^d$\citet{B:97}, $^e$\citet{L:05},\\
$^f$\cite{S:02}, $^g$\cite{C:99.2}, $^h$\cite{P:92}\\}
\end{center}
\end{table}

For the further analysis, we included two additional data from literature 
sources. \citet{B:97} determined the ratio of CN-strong to CN-weak stars for 
stars on the RGB in 47\,Tuc. He distinguished between RGB stars below and 
above the RGB bump and found very similar values of 1.9 and 1.8, respectively.
For this work, we adopted the value of 1.9. \citet{P:92} and \citet{L:05},
who found the r-parameter in the cluster M\,71 for stars on the lower RGB to 
be 0.8 and 0.69, respectively. We adopted the more recent result by 
\citet{L:05}. The measurements of the number ratio of CN-strong stars on the 
upper RGB of M\,71 range between 0.3 \citep{P:92}, 0.63 \citep{C:99.2}, and 1.0 
\citep{L:05}. The average value is 0.64, similar to those found 
on the lower RGB. Nevertheless, we have to keep in mind that the additional 
r-parameters are based on observations obtained with a different instrument 
and different index definitions.

In the upper part of Table~\ref{tab:ratio} an overview of the number of stars 
identified as CN-strong and CN-weak is given. In the third column the 
r-parameters for the lower RGB of our clusters and M\,71 and 47\,Tuc are 
listed. The r-parameters range from $0.0$ for Ter\,7 to $2.00$ for NGC\,5286.
The uncertainties vary from $0.07$ for Pal\,12 to $1.73$ for NGC\,5286. The 
large uncertainty for NGC\,5286 is due to the small sample size. If one would
divide the stars of M\,15 into CN-weak and CN-strong according to 
Fig.~\ref{CNCH} its r-parameter would be 0.6 (given in brackets in 
Table~\ref{tab:ratio}). For those clusters that were part 
of previous studies the r-parameters for the upper RGB are given in the last 
column. We find for two out of the three clusters of our sample, for which RGB
studies exist, a good agreement of the number ratios found on the SGB with 
those on the RGB. The values for NGC\,362 differ by a factor of 2. The reason 
for this remains unclear and requires the repetition of the measurement on the 
RGB.

\section{CN-CH - anticorrelation}\label{MWGC:anti}
In many clusters the bimodal distribution in CN is accompanied by an 
anticorrelation of CN and CH. A summary on this can be found in e.g., 
\citet[][]{K:94}. As these abundance patterns are similar to those expected by
the nucleosynthesis of material in the CNO cycle they have been attributed to 
a dredge-up of processed material to the stellar surfaces. In the meantime 
CN-CH anticorrelations have been found to be very common for clusters with a 
bimodal distribution in CN \citep[see, e.g., the recent review paper 
by][]{G/S/C:04}.

\begin{figure}[t!]
\centering
\psfig{figure=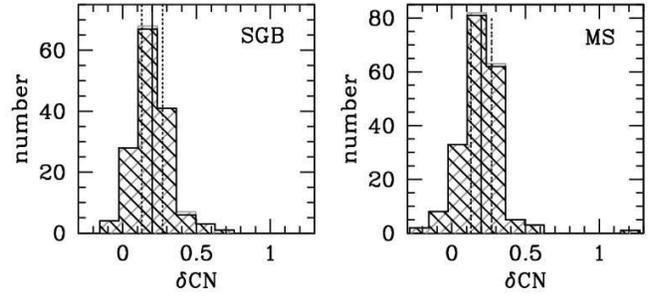,width=8cm,bbllx=25mm,bblly=110mm,bburx=200mm,bbury=210mm}
\caption{\label{fig:SGBMS} The combined distributions of the CN excess 
parameters measured for stars on the SGB and MS.  The grey histogram 
comprises all eight clusters. In the black histogram Pal\,12 and Ter\,7 are 
not included. The solid lines indicate the median values of the distributions. 
The dashed lines indicate the selection limits for CN-strong and CN-weak 
stars. Stars with \dcn\ smaller than the position of the first dashed line are
considered as CN-weak, stars with \dcn\ larger than the position of the second
dashed line as CN-strong.}
\end{figure}

In order to examine possible CN-CH anticorrelations we used the distinction 
criteria between CN-strong and CN-weak RGB stars as described in 
Section~\ref{sec:CN}. Although no clear bimodality in CN absorption strength 
was detected on the SGB and MS, we observe a scatter in CN larger than 
expected from measurement errors alone in all evolutionary states.

Since CN dichotomies have been detected before on the MS on M13 \citep{B/H:04}, 47\,Tuc 
\citep{H/S/G:03}, and M\,71 \citep{C:99.2} it is quite conceivable that 
abundance variations among the less evolved stars exist in our sample as well.
At the precision of our measurements, however, the signal might be simply too 
weak due to the higher temperatures and/or low metallicities, which inhibit the
formation efficiency of the CN molecule.
Nevertheless, in order to check for anticorrelations, we determined the CN 
excess parameter for the SGB and MS stars analogously to the RGB stars.
The resulting \dcn\ distributions are shown in Fig.~\ref{fig:SGBMS}.
In analogy to the RGB analysis we neglected the Sgr clusters Pal\,12 and 
Ter\,7. The median \dcn\ values were found to be 0.20 both for the SGB and MS. 
The standard deviation is 0.08 in both cases.
We considered those stars with \dcn\ higher than $1\sigma$ above and below the 
median as CN-strong and CN-weak, i.e. CN-strong: \dcn$ > {\rm median}+\sigma$;
CN-weak: \dcn $< {\rm median}-\sigma$.

\begin{figure*}[]
\centering
\begin{tabular}[]{c c}
\psfig{figure=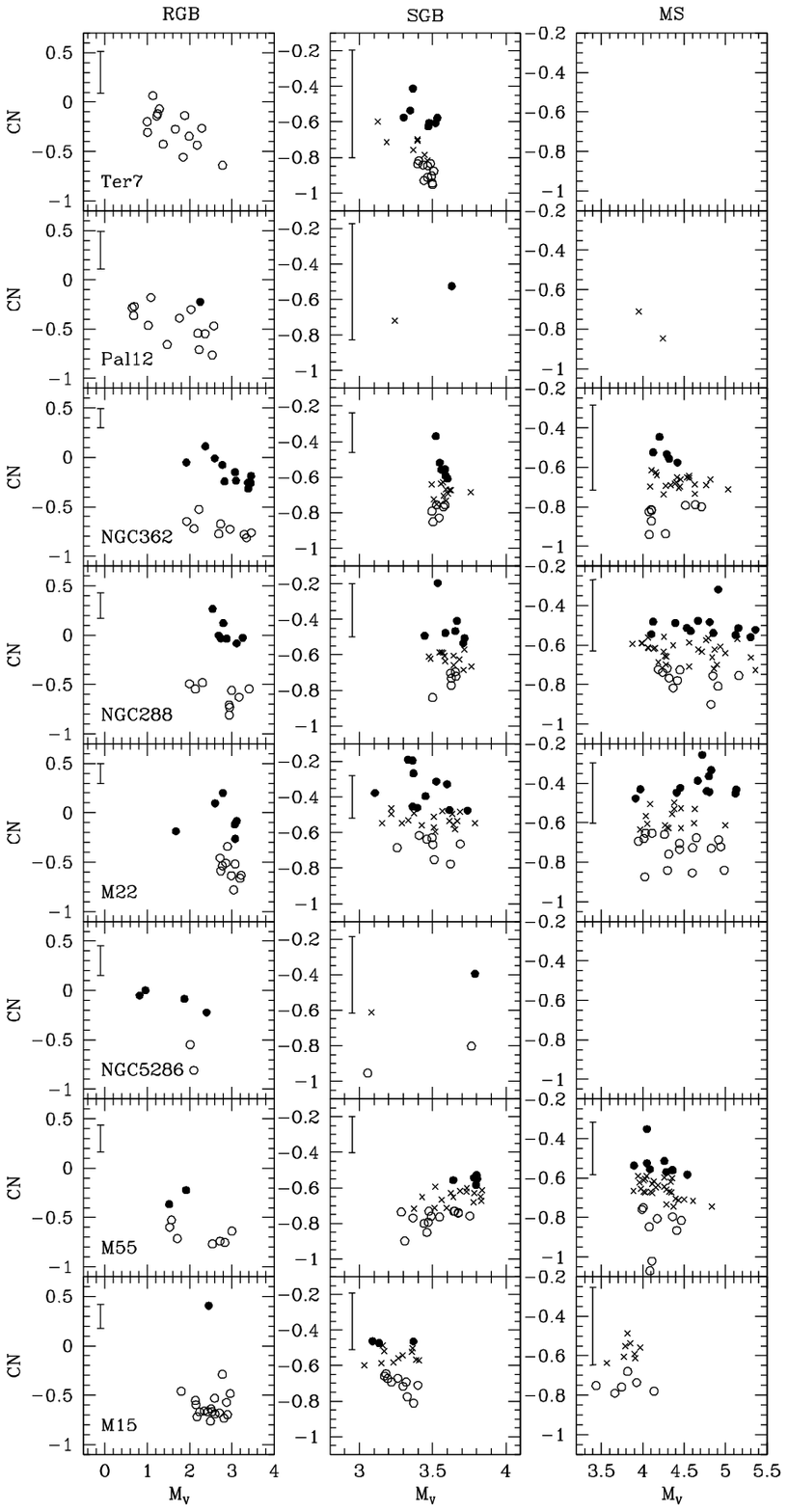,height=18cm,bbllx=25mm,bblly=60mm,bburx=114mm,bbury=246mm}&
\psfig{figure=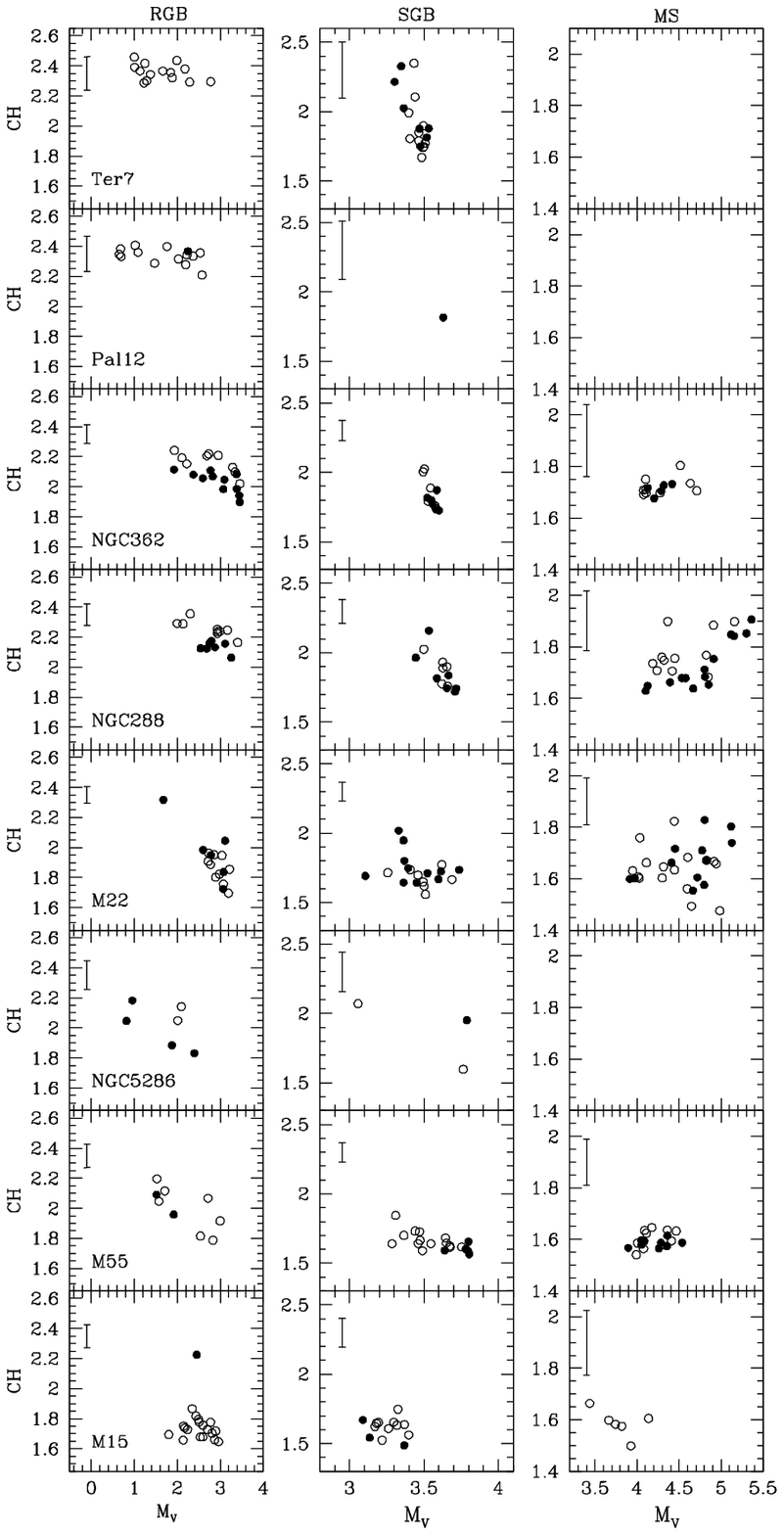,height=18cm,bbllx=110mm,bblly=60mm,bburx=550mm,bbury=246mm}
\end{tabular}
\caption{\label{anti} The CN vs. \mv\ and CH vs. \mv\ diagrams for the eight 
clusters in our sample. We differentiate between RGB, SGB, and MS stars for 
all clusters. The CN-strong and CN-weak stars are marked by filled and open 
symbols, respectively. SGB and MS stars with intermediate \dcn\ strength are 
plotted as crosses. For reasons of clarity these stars are only plotted in the 
CN vs. \mv\ diagrams. The median error of the measurements are given in the 
upper left corners of each panel. We do not plot the errors for SGB and MS stars in 
Pal\,12 as they exceed the limit of the diagrams.}
\end{figure*}

A comparison of CN vs. \mv\ and CH vs. \mv\ is shown in Fig.~\ref{anti}. We 
differentiate between RGB, SGB, and MS stars for all clusters. Stars with 
stronger and weaker CN absorption band features are highlighted by filled and 
open symbols, respectively. SGB and MS stars with intermediate \dcn\ strength 
are plotted as crosses in the CN vs. \mv\ diagrams only. A bimodal 
distribution in CH is not detected for any of the clusters. Note that even for
NGC\,288 and NGC\,362, which showed the strongest dichotomy in CN, we do not 
observe a bimodality in CH.
However, the CN-strong RGB stars of these two GCs clearly have smaller CH 
indices than the CN-weak RGB stars of similar \mv . This is not seen for the
other clusters, except maybe for NGC\,5286. In case of M\,22, larger 
uncertainties due the significant differential reddening \citep{R/H:99} might
dilute a possible CN-CH anticorrelation.
In the very metal-poor cluster, M\,15, one RGB star with high CN absorption 
bands was identified, which also seems to be quite rich in CH. This CN- and 
CH-strong star in M\,15 stands out from the rest of the datapoints by more 
than 1 in \dcn . Since this star lies slightly off the RGB we suggest 
that this star is not a cluster member (although it has the right radial
velocity).

Moving from the RGB to the SGB and the MS, the CN-CH anticorrelation 
still is visible for NGC\,288 and NGC\,362. Due to the smaller signal to noise
ratio, it is less pronounced but on average the more CN-strong stars are more 
CH-weak. For the other clusters, no clear statement can be made.

We conclude that in case of clearly bimodal clusters like NGC\,288 and 
NGC\,362 the differences in the band strengths and the CN/CH anticorrelation
do exist among stars of all evolutionary states. Deep mixing is believed to 
set in at the level of the RGB bump and does not take place in stars on the 
lower RGB, SGB, and MS. Furthermore low-mass MS stars burn hydrogen in their 
cores only. Thus the observed patterns can not be caused by the transport of 
CNO cycle processed material from the interior to the stellar surfaces.
We can therefore rule out evolutionary effects within the stars as the origin 
of the observed anticorrelation.

\begin{table*}[]
\caption{\label{tab2}Global parameters of globular clusters of our sample. All
sources for the parameters are listed in the footnotes.}
\begin{center}
\begin{tabular}[l]{cccccccccp{1.2cm}ccccc}
\hline\hline
Cluster      & $D_{GC}^{1a}$ & $M_V^{a}$&HBR$^{2a}$&[Fe/H]$^{a}$&$c^{3a}$&$e^{4a}$&$r_{core}^{5a}$ &$r_{tidal}^{6a}$& $\mu_{0,V}^{7a}$  & $\sigma^{8b}$ & (M/L)$^b$  & age$^{c,d}$ & class$^f$\\
              &   (kpc)    & (mag)    & 	  & (dex)  & 	     & 	        & (pc)	&(pc)  & (mag/$\arcsec^2$) &   (km/s)    &     &   & (Gyr)   \\
 \hline
NGC\,288      &   12.0        &  -6.74  &  0.98  &  -1.24  &  0.96   & 0.09$^e$ & 3.64  &  33.12  &  19.95  &   2.9  &  3.0  &  11.3   &OH\\
NGC\,362      &    9.4        &  -8.41  & -0.87  &  -1.16  &  1.94   &  0.01    & 0.47  &  39.83  &  14.88  &   6.4  &  1.1  &   8.7   &YH\\
NGC\,5286     &    8.4        &  -8.61  &  0.80  &  -1.67  &  1.46   &  0.12    & 0.93  &  26.75  &  16.07  &   8.0  &  2.1  &  NA     &OH\\
 M\,22	      &   4.9	      &  -8.50  &  0.91  &  -1.64  &  1.31   &  0.14    & 1.32  &  26.97  &  17.32  &   9.0  &  3.3  &  12.3   &OH\\
Terzan\,7     &   16.0        &  -5.05  & -1.00  &  -0.58  &  1.08   &   NA     & 4.12  &  49.06  &  20.65  &   NA   &   NA  &   7.4   &SG\\
M\,55	      &    3.9        &  -7.55  &  0.87  &  -1.81  &  0.76   &  0.02    & 4.36  &  25.10  &  19.13  &   4.9  &  3.4  &  12.3   &OH\\
M\,15	      &   10.4        &  -9.17  &  0.67  &  -2.26  &  2.50   &  0.05    & 0.21  &  64.42  &  14.21  &  12.0  &  2.2  &  12.3   &YH\\
Palomar\,12   &   15.9        &  -4.48  & -1.00  &  -0.94  &  1.94   &   NA     & 1.11  &  96.78  &  20.59  &   NA   &   NA  &   6.4   &SG\\
\hline
47\,Tuc        &  7.4        & -9.42  &  -0.99  &  -0.76   &  2.03   &   0.09	& 0.52  &  56.1   & 14.43   &  11.5  &  2.0  &   10.7  & BD\\
M\,71          &  6.7        & -5.60  &  -1.00  &  -0.73   &  1.15   &   0.00	& 0.73  &  10.43  & 19.22   &	2.3  &  1.1  &   10.2  & BD\\
\hline
NGC\,1904      &  18.8	     & -7.86  &   0.89  &  -1.57   &  1.72   &   0.01	& 0.60  &  31.3   & 16.23   &	5.4  &  2.2   &  11.7  & OH\\
NGC\,2808      &  11.1       & -9.39  &  -0.49  &  -1.15   &  1.77   &   0.12	& 0.73  &  43.42  & 15.17   &  13.4  &  2.4   &   9.3  & OH\\
NGC\,3201      &   8.9       & -7.46  &   0.08  &  -1.58   &  1.30   &   0.12	& 2.08  &  41.38  & 18.77   &	5.2  &  4.1   &  11.3  & YH\\
NGC\,5272      &  12.2       & -8.93  &   0.08  &  -1.57   &  1.84   &   0.04	& 1.66  &  115.5  & 16.34   &	5.6  &  1.2   &  11.3  & YH\\
NGC\,5904      &   6.2       & -8.81  &   0.31  &  -1.27   &  1.83   &   0.14	& 0.92  &  61.96  & 16.05   &	5.7  &  1.4   &  10.9  & OH\\
NGC\,6121      &   5.9       & -7.20  &  -0.06  &  -1.20   &  1.59   &   0.00	& 0.53  &  20.79  & 17.88   &	4.2  &  2.6   &  11.7  & OH\\
NGC\,6171      &   3.3       & -7.13  &  -0.73  &  -1.04   &  1.51   &   0.02	& 1.01  &  32.47  & 18.84   &	4.1  &  3.9   &  11.7  & OH\\
NGC\,6205      &   8.7       & -8.70  &   0.97  &  -1.54   &  1.51   &   0.11	& 1.75  &   56.4  & 16.80   &	7.1  &  2.2   &  11.9  & OH\\
NGC\,6254      &   4.6       & -7.48  &   0.98  &  -1.52   &  1.40   &   0.00	& 1.10  &  27.49  & 17.69   &	6.6  &  3.5   &  11.8  & OH\\
NGC\,6637      &   1.9       & -7.64  &  -1.00  &  -0.70   &  1.39   &   0.01	& 0.90  &  22.10  & 16.83   &	NA   &   NA   &  10.6  & BD\\
NGC\,6752      &   5.2       & -7.73  &   1.00  &  -1.56   &  2.50   &   0.04	& 0.20  &  64.39  & 15.20   &	4.5  &  1.1   &  12.2  & OH\\
NGC\,6934      &  12.8       & -7.46  &   0.25  &  -1.54   &  1.53   &   0.01	& 1.14  &  38.23  & 17.26   &	5.1  &  2.5   &   9.6  & YH\\
NGC\,7089      &  10.4       & -9.02  &   0.96  &  -1.62   &  1.80   &   0.11	& 1.14  &  71.75  & 15.92   &	8.2  &  1.9   &  NA    & OH\\
NGC\,7006      &  38.8       & -7.68  &  -0.28  &  -1.63   &  1.42   &   0.01   & 2.90  &  76.54  & 18.50   &	NA   &  NA  &  NA    & YH\\
\hline\hline
\end{tabular}
\end{center}
\vspace{-0.2cm}
\footnotesize{$^1$distance from Galactic center,
	      $^2$horizontal branch ratio: $\rm{HBR=(B-R)/(B+V+R)}$,
	      $^3$concentration,
	      $^4$ellipticity $e=1-(b/a)$,
	      $^5$core radius,
	      $^6$tidal radius,
	      $^7$central surface brightness,
	      $^8$central velocity dispersion\\
	      $^a$\citet{H:96}
	      $^b$\citet{P/M:93}
	      $^c$\citet{R/S:99}
	      $^d$\citet{B/C:98}
	      $^e$\citet{F/F:82}
	      $^f$\citet{M/VDB:05}\\}
\end{table*}

\section{Trends with cluster parameters}\label{MWGC:trends}
\begin{figure*}[]
\centering
\psfig{figure=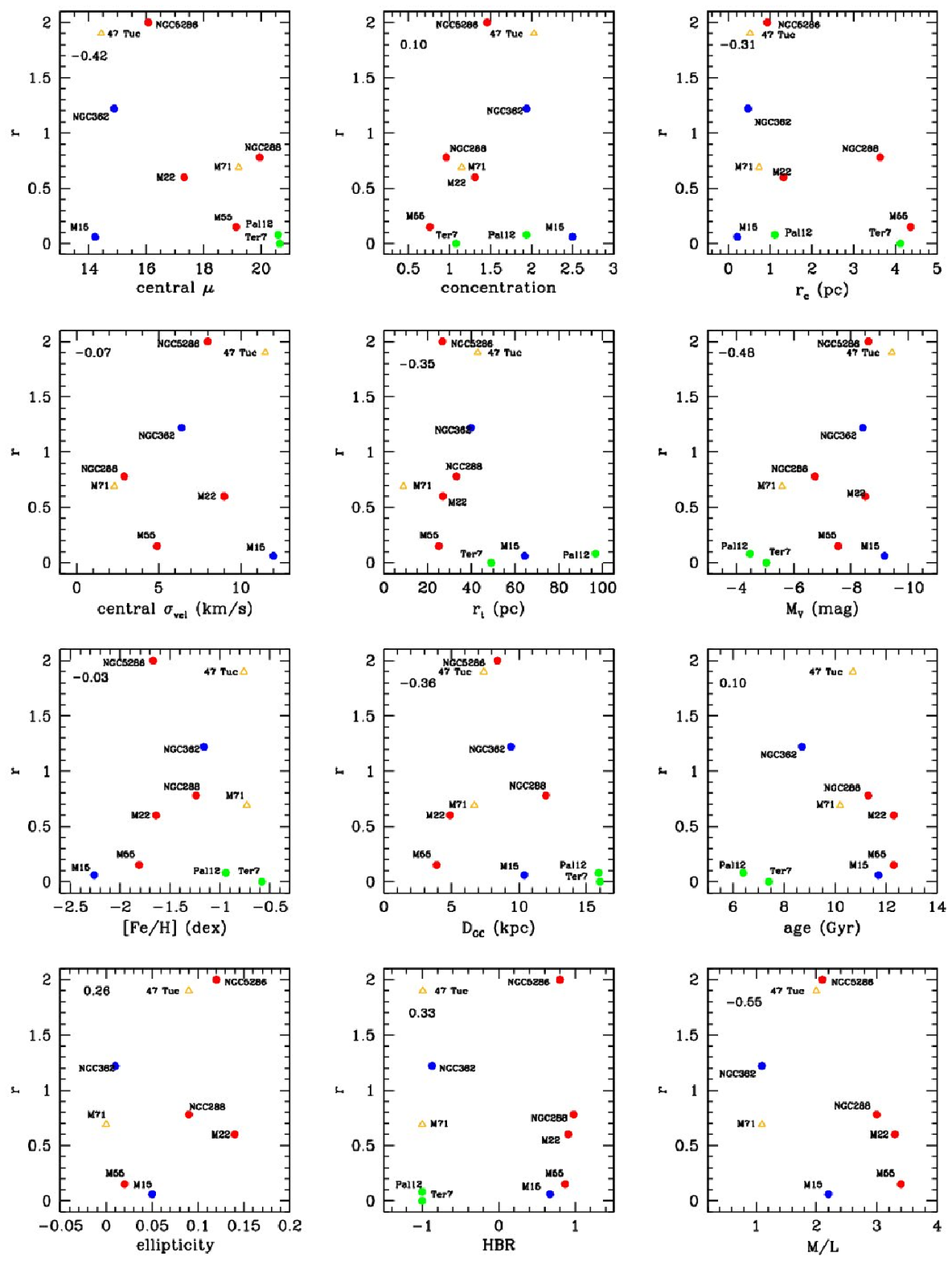,height=23cm,bbllx=30mm,bblly=20mm,bburx=150mm,bbury=200mm}
\caption{\label{trends} The number ratio of CN-strong to CN-weak stars (r-parameter) vs. various cluster 
parameters (see Tab.~\ref{tab2}). Our targets are indicated by the filled circles. The two results 
taken from the literature are marked by open triangles. Red, blue, yellow and
green colours indicate OH, YH, BD and Sgr GCs, respectively. In the upper left 
corner the calculated Spearman rank correlation coefficient is given.}
\end{figure*}

In order to explore possible correlations of the CN distribution with global 
parameters of the globular clusters we combine our observations to quantities 
available in the literature. A similar analysis was done before by, e.g., 
\cite{N:87}, \cite{S/M:90}, \cite{S:02} and \cite{H/S/G:03b}. However their 
studies were based on compilations of upper RGB star measurements in various 
clusters from different sources and therefore different techniques. We now 
provide a sample that is based on a very homogeneous data set of eight star 
clusters. The cluster quantities were selected from the 2003 version of the 
McMaster \citep{H:96} and \citet{P/M:93} globular cluster 
catalogs\footnote{\texttt{http://coihue.rutgers.edu/$\sim$andresj/gccat.html}}.
As no ellipticity is listed in these catalogs for NGC\,288, we adopted the 
value given by \citet{F/F:82}. The age estimates were adopted from 
\citet{R/S:99} and \citet{B/C:98}. Moreover, we adopted the subdivision of our
globular clusters into objects belonging to different Galactic components 
(namely OH, YH, BD, and those accreted from the Sgr dSph (SG)) from 
\citet{M/VDB:05}. Table~\ref{tab2} gives an overview of the extracted 
parameters.

In order to quantify the statistical significance of possible correlations 
between the number ratio of CN-strong stars with various structural parameters
we computed for each parameter the Spearman coefficient of rank correlation, 
$r_s$. This correlation coefficient is a technique that can be used to 
characterize the strength and direction of a relationship of two random 
variables.
The values of $r_s$ lie between $+1$ and $-1$, the extremes where the rank
sequences completely coincide and are completely opposite, respectively.
For the clusters in our sample we do not find a clear correlation between the 
majority of the cluster parameters and the percentage of CN-strong stars 
(Fig.~\ref{trends}).

\citet{N:87} observed a correlation between the percentage of CN-rich stars 
and the apparent flattening of the individual clusters, which he proposed to be
associated with the clusters' rotation. He suggested that the high systematic 
cluster rotation is linked, via exchange of angular momentum, to a higher 
initial angular momentum of the individual stars. Within giants the rotation 
may drive circulation currents that are capable of cycling the material in the
envelope through the interior hydrogen-burning shell where the CNO process is 
active \citep{S/M:79}. Consequently a larger percentage of CN-strong stars is 
expected to be observed in clusters with larger mean stellar rotation 
velocities and thus larger overall cluster rotational velocities and hence 
possibly larger ellipticities. Since there is little information on cluster 
rotation for the globulars in our sample, we use ellipticity as a proxy for 
rotation. This correlation was confirmed by \citet{S/M:90} and \citet{S:02}.
The computed Spearman rank coefficient of 0.26 suggests that the number ratio
of CN-strong stars is mostly independent of the cluster ellipticity. We 
conclude that the effect proposed by \citet{S/M:79} is probably not as 
relevant as thought so far.
\begin{table}[]
\caption{\label{tab5} Calculated Spearman rank correlation coefficients.}
\begin{center}
\begin{tabular}[t]{cccc}
\hline\hline
              	& all clusters  & alternative value & without \\
		&		&  for M\,15        & NGC5286 \\
 \hline
$\mu_{0,V}$   	&  $-0.42$  &  $-0.58$ 	& $-0.38$	\\
$c$     	&  $\;\;\; 0.10$  &  $\;\;\; 0.24$ 	& $\;\;\; 0.07$	\\
$r_{core}$    	&  $-0.31$  &  $-0.48$ 	& $-0.30$	\\
$\sigma$   	&  $-0.03$  &  $\;\;\; 0.05$ 	& $-0.14$	\\
$r_{tidal}$   	&  $-0.35$  &  $-0.27$ 	& $-0.27$	\\
$M_V$   	&  $-0.48$  &  $-0.62$ 	& $-0.43$	\\
${\rm [Fe/H]}$  &  $-0.03$  &  $-0.12$ 	& $\;\;\; 0.10$	\\
$D_{GC}$   	&  $-0.36$  &  $-0.31$ 	& $-0.38$	\\
 age   		&  $\;\;\; 0.10$  &  $\;\;\; 0.18$ 	& $\;\;\; 0.10$	\\
$e$		&  $\;\;\; 0.26$  &  $\;\;\; 0.29$ 	& $\;\;\; 0.07$	\\
 HBR		&  $\;\;\; 0.33$  &  $\;\;\; 0.36$ 	& $\;\;\; 0.32$	\\
 M/L		&  $-0.55$  &  $-0.62$ 	& $-0.57$	\\
\hline\hline
\end{tabular}
\end{center}
\end{table}

Another correlation detected by \citet{S/M:90} is between the r-parameter and
the central velocity dispersion. Our analysis reveals $r_s = -0.07$, which 
makes such a correlation rather unlikely. Furthermore, \citet{S/M:90} found 
the largest percentages of CN-strong stars to be restricted to the more 
luminous/massive clusters. They suggest an inter-cluster self-pollution 
scenario as a possible origin. Due to the higher binding energies in more 
massive clusters, the ability to retain enriched ejecta of massive and 
intermediate-mass stars is expected to be higher than in lower mass clusters.
Our cluster sample supports the correlation with the total absolute magnitude ($M_V$).
The calculated Spearman coefficient of $r_s=-0.48$ is actually among the highest 
found in our analysis.

We furthermore determined the Spearman rank correlation coefficients 
using the alternative higher number ratio of M\,15 (see Sect.~\ref{sec:CNvsCH}).
Although most of the changes are small some correlations show a higher significance,
in particular for $M_V$ with $r_s=-0.62$.
Since the results for NGC\,5286 suffer from small number statistics and thus a large
error in $r_s$ we decided to also recalculate the correlation coefficients by neglecting
this cluster (using the original value for M\,15).
The resulting values are very comparable to those considering all clusters.
An overview of the computed Spearman rank coefficients is given in Tab.~\ref{tab5}.

\begin{figure*}[t!]
\centering
\begin{tabular}{cc}
\psfig{figure=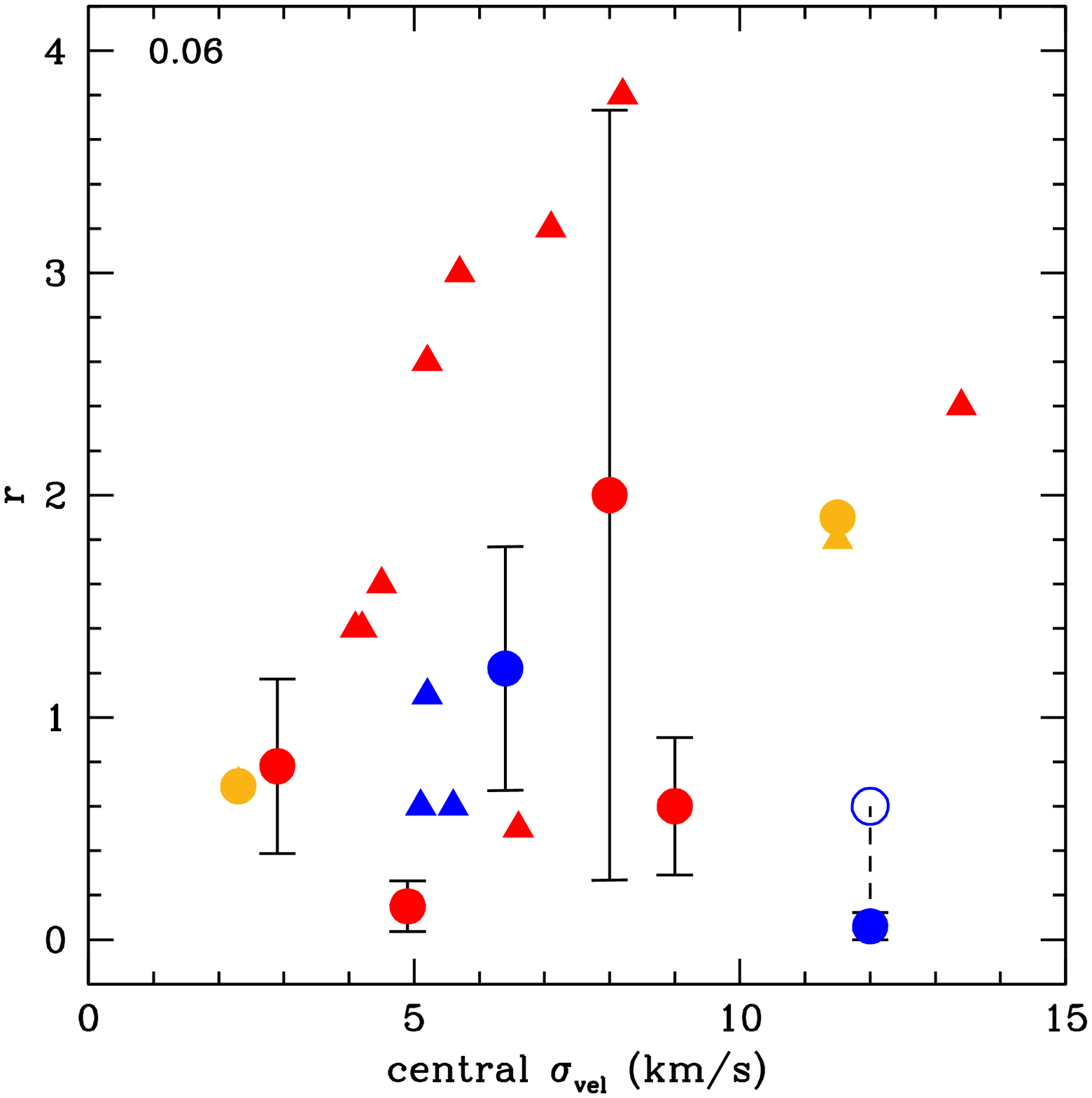,width=7.8cm,bbllx=25mm,bblly=60mm,bburx=200mm,bbury=246mm}&
\hskip 0.3cm
\vspace{0.5cm}
\psfig{figure=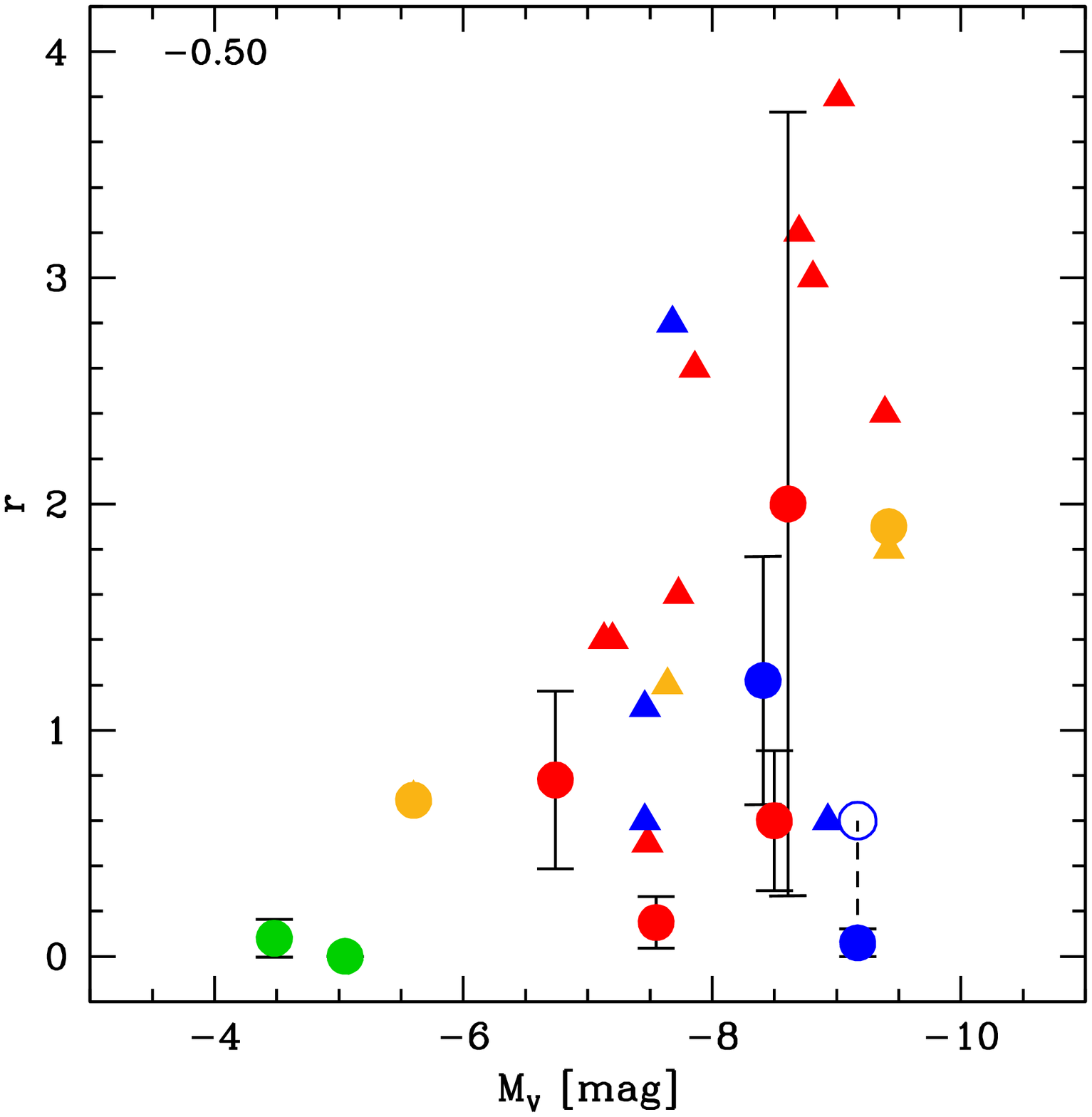,width=7.8cm,bbllx=25mm,bblly=60mm,bburx=200mm,bbury=246mm}\\
\psfig{figure=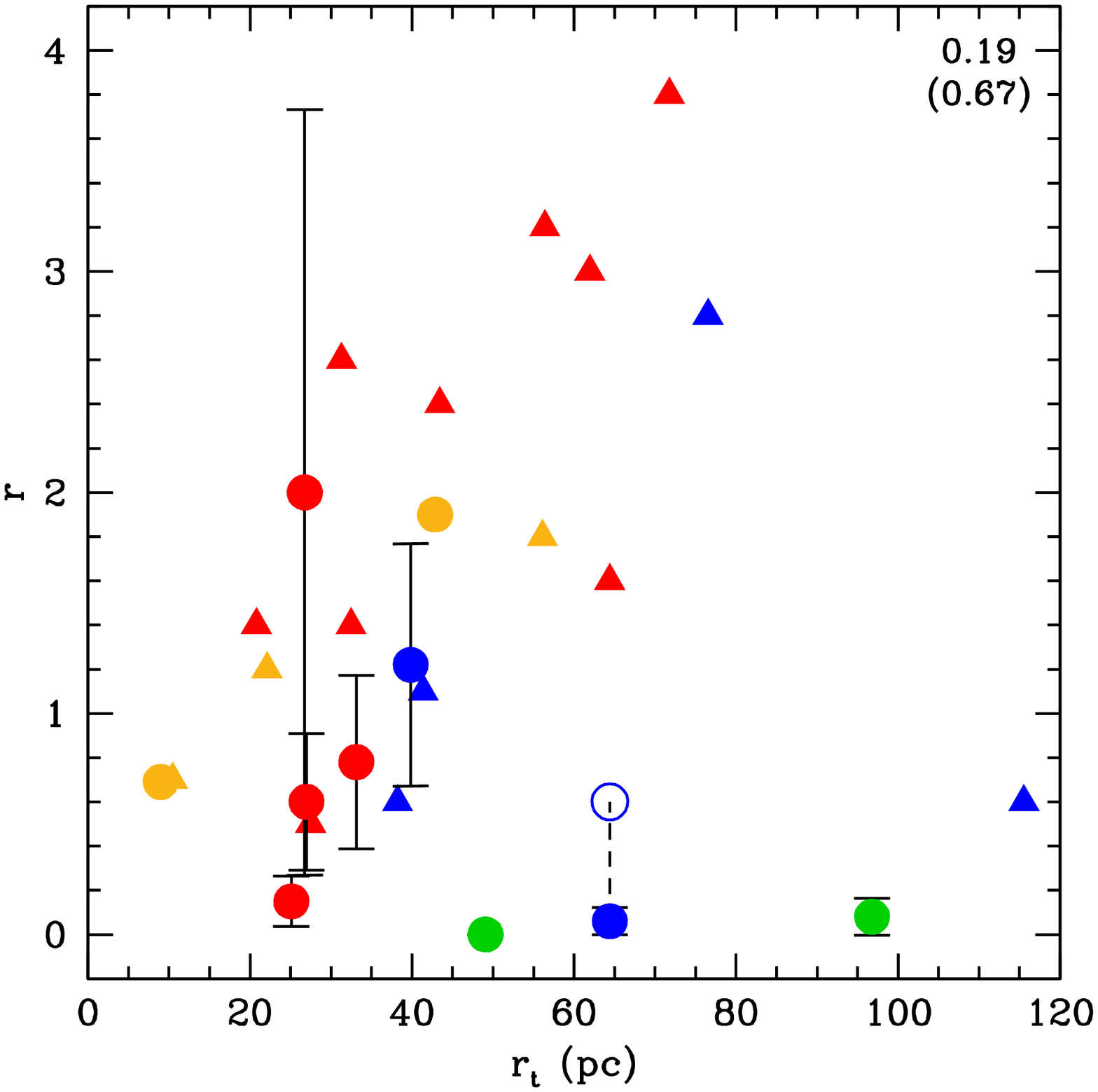,width=7.8cm,bbllx=25mm,bblly=60mm,bburx=200mm,bbury=246mm}&
\hskip 0.3cm
\psfig{figure=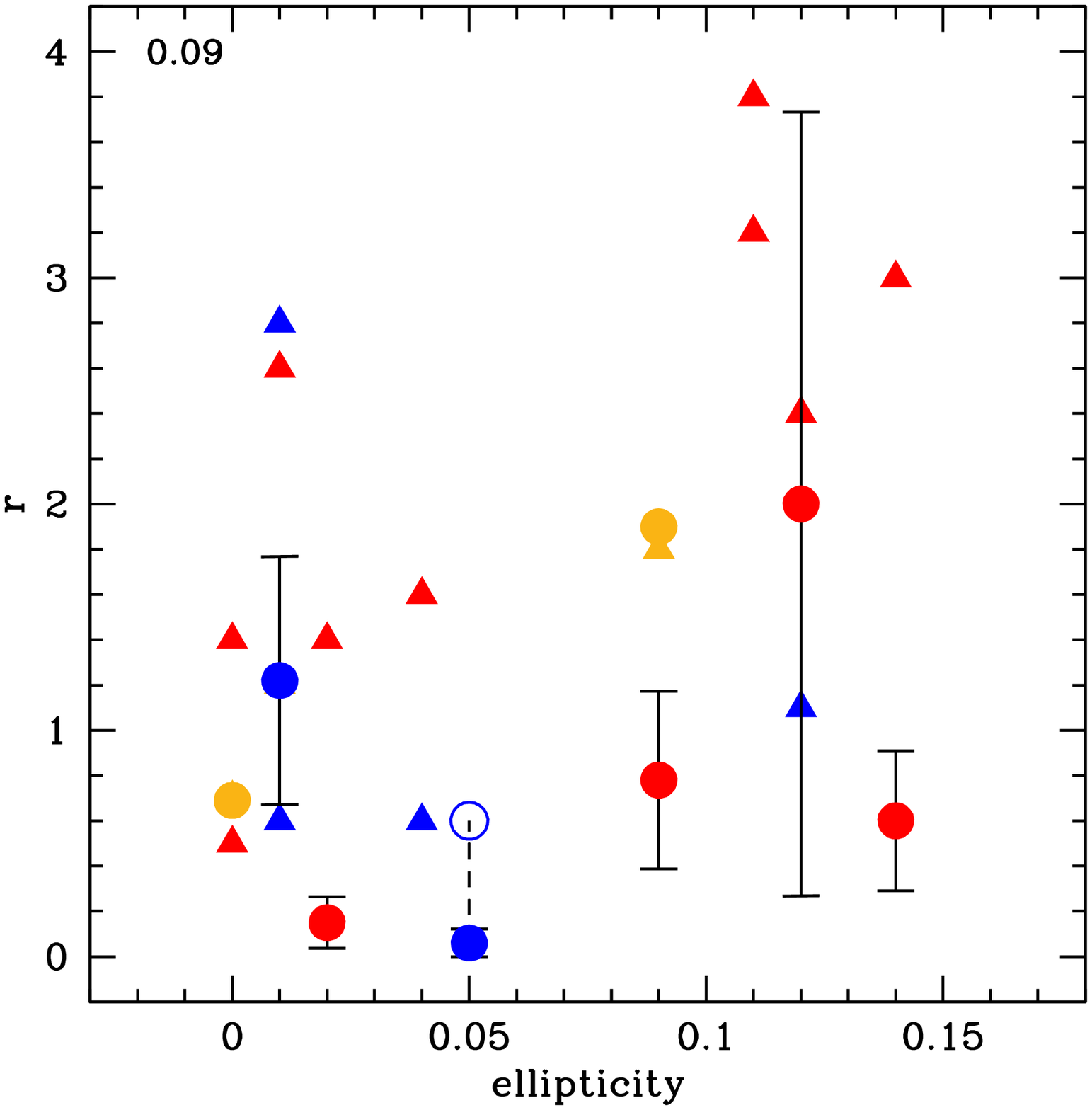,width=7.8cm,bbllx=25mm,bblly=60mm,bburx=200mm,bbury=246mm}\\
\end{tabular}
\caption{\label{smith} Plot of the r-parameter vs. those globular cluster 
parameters that showed a promising correlation in Fig.~\ref{trends} or in previous studies by e.g., \citet{S:02}, such as 
ellipticity, central velocity dispersion ($\sigma_{\rm vel}$), absolute brightness ($M_V$) and tidal radius ${\rm r_t}$. We also 
included the results by \citet{S:02} (triangles). The results of our study are
plotted as filled circles. In this figure we furthermore differentiate between
the different MW cluster populations. Old and young halo clusters are colored 
in red and blue, respectively. Bulge and disk clusters are plotted in yellow 
and accreted clusters from dwarf galaxies such as the Sagittarius dSph in 
green. The open blue circle indicates the alternative value of the r-parameter for M\,15.}
\end{figure*}

In order to perform a more statistically complete investigation we combined 
our results with those by \citet{S:02} and \cite{H/S/G:03b}. In 
Section~\ref{sec:CN} we have seen that for the majority of the studied 
clusters the r-parameter on the upper RGB is consistent with those on the 
lower RGB. We are thus confident that we may combine our results with those from the 
literature. Nevertheless we keep in mind that this leads to a more 
heterogeneous sample, since values of different evolutionary states and 
different measurements are combined.

For most parameters the lack of any clear trends is confirmed. In particular,
the inclusion of our results with those listed in \citet{S:02} and 
\citet{H/S/G:03b} further confirms the lack of a correlation between cluster 
ellipticity $\epsilon$ and the number ratio of CN-strong stars.
We observe a large scatter in Fig.~\ref{smith} (lower right panel). It can,
however, not be ruled out that some clusters with low $\epsilon$ and high
r values are actually more elliptical but appear round due to projection
effects. This would dilute a possible correlation. GCs with high $\epsilon$
and low r values would then be clear outliers.

We see a possible connection between the r-parameter and the tidal radius
(Fig.~\ref{smith}, lower left panel). Clusters with larger tidal radii seem to
possess a larger percentage of CN-strong stars. Interestingly, those clusters 
that do not follow this trend are those that are thought to be linked to the 
Sgr dSph (Palomar\,12 and Terzan\,7) as well as the very metal-poor clusters 
M15 and NGC\,5272, which belong to the young halo GCs. We computed the Spearman
coefficient including and excluding these cluster. The resulting values are 
0.19 and 0.67, respectively. This is an interesting finding, since \cite{Z:93}
postulated that the young halo population of globular clusters was 
predominantly formed by accretion of extragalactic objects.
We therefore put forward the hypothesis that, among other parameters, 
environmental differences due to different cluster formation sites may 
influence the today observed abundance patterns. \cite{C:06} showed that apart
from differences in the environmental properties during the time of formation 
also differences in the evolution of clusters have probably influenced the 
light element abundance ratios.
Using a set of high resolution spectroscopic abundance measurements he found 
that clusters with larger orbital semi major axes, i.e., extended orbits and 
revolution periods, exhibit a larger amount of inhomogeneities. From this he 
concluded that for clusters on orbits reaching large Galactocentric distances 
the lack of disturbance by the Galactic disk helps to retain pre-enriched 
material. In contrast, clusters close to the Galactic centre might have
suffered early and frequent disk/bulge shocks that enforced rapid gas loss and
prohibited the formation of a second enriched subpopulation. Those clusters
also show smaller tidal radii due to the even stronger tidal forces towards
the centre of the Galaxy.

In the upper right panel of Fig.~\ref{smith} we plotted the r-parameter as a 
function of the absolute magnitude, representing the present-day cluster mass.
It seems that the maximum number ratio of CN-strong to CN-weak stars increases
with increasing $M_V$ \citep[cf.][]{S:02}. Only the brightest clusters have formed CN-strong 
stars. This supports the idea that the more massive objects can more 
efficiently retain processed material ejected from evolved stars.

The possible CN-bimodality of M\,15 as described in Sect.~\ref{sec:CNvsCH} and 
shown in Fig.~\ref{CNCH} increases the r-parameter of this cluster to 0.6.
As a consequence the correlations of r with absolute magnitude, $M_V$ and tidal radius, $r_t$ become slightly more
significant with Spearman rank values of $r_s=-0.56$ and $-0.68$,
respectively. 
The low correlations with central velocity dispersion and ellipticity, however, remain nearly unchanged.
More accurate CN/CH index measurements of this very metal-poor
cluster are needed to confirm these findings.

\section{Summary and conclusions}\label{MWGC:sum}
We analyzed the absorption bands of the CN and CH molecule in eight Galactic 
globular clusters via line index measurements. In each cluster, stars of 
various evolutionary stages were studied, from the lower RGB and SGB to the 
upper MS. Our sample comprises clusters belonging to different Milky Way 
components, e.g., young and old halo. In particular, two of our studied 
objects are associated with a disrupting Galactic companion, the Sagittarius
dwarf spheroidal (Sgr dSph). We could show that the majority of the studied 
clusters shows significant CN/CH variations at the base of the RGB. For the
two most prominent CN-bimodal GCs, NGC\,288 and NGC\,362, CN anticorrelates 
with CH. A weak signal for a CN/CH anticorrelation was detected also in the 
least evolved stars in these clusters. From this we conclude that purely 
evolutionary effects within the stellar interior cannot be the main driver of 
the observed abundance patterns. Our findings therefore favor a scenario in 
which a certain fraction of most clusters was formed out of material that was
enriched or polluted by ejecta of a prior generation of massive stars. 
In fact, the existence of star-to-star variations among those slightly evolved 
stars favors self-enrichment as the probable origin. One possible explanation
could be that the nowadays observed stars in globular clusters formed from 
protocluster material that was to some degree inhomogeneously enriched in 
light elements. Such a pollution might have originated from ejecta of a prior 
generation of massive and therefore fast evolving stars, either belonging to
the cluster itself or to the field population of a larger (dwarf sized) 
galaxy in which the cluster was embedded \citep[e.g.,][]{B/C:07}. Possible 
candidates for the polluters discussed in the literature are massive AGB 
stars \citep[e.g.,][]{C/DC:81,V/DA:01} and more recently fast rotating massive
stars \citep{D/M:07}. AGBs eject material via slow winds that are processed 
through the hot CNO cycle but are not enriched in iron. Fast rotating massive
stars loose large amounts of material through slow winds, which are also 
enriched in H-burning products.

For the clusters NGC\,288 and NGC\,362 we found a clear bimodal distribution 
in CN with similar numbers of CN-strong and CN-weak stars.
As the two clusters are a second-parameter pair, 
we conclude that the horizontal branch morphology is not correlated with this 
phenomenon. A possible explanation for such a pronounced dichotomy is given 
by a prolonged star formation in these globular clusters. The second, enriched
stellar population formed well after the first generation had expelled and
homogeneously distributed their AGB ejecta. The existence of such multiple 
stellar populations within globular clusters is further supported by the 
recent discoveries of complex CMD morphologies (e.g., multiple SGBs and MSs 
with age spreads) in some massive objects \citep{B/P:04, P/B:07}.

The two probable former Sgr dSph clusters (Terzan\,7 and Palomar\,12) do not 
exhibit any CN-strong stars. They are the most metal-rich clusters in our 
sample and therefore the double metal molecule CN should be easy to detect.
We conclude that these clusters might lack stars with strong CN 
absorption. Our results suggest that the accreted Sgr globular clusters might 
be more chemically homogeneous than those native to the Milky Way. This is
supported by the abundance analysis of 21 elements for four Sgr stars by 
\cite{C:04}, who do not find a significant star-to-star scatter. Probably 
environmental conditions during the formation of the clusters played a major 
role for the observed abundance pattern. 
However, we point out that all existing studies suffer from small number statistics.
Thus it can not be ruled out that CN-rich stars are simply missed in the sampling
\citep[see the priliminary results by][]{B/M/S:07}.
For further conclusions a thorough investigation of the abundance patterns other
probable Sgr clusters (M\,54, Arp\,2, Ter\,7, Ter\,8, and Pal\,12) is 
desirable.

In order to search for possible drivers for the abundance anomalies we studied
the ratio of CN-strong/CN-weak stars as a function of various cluster
parameters. We do not confirm the correlation with the cluster ellipticity 
that was observed before \citep{N:87}. Our study therefore does not support 
cluster rotation and the associated enhanced deep mixing \citep{S/M:79} 
as a main source of the production of CN-strong stars.
Although we hardly see correlations of the number ratio of CN-strong stars 
with the majority of cluster parameters, some dependencies do seem to exist. We 
find evidence for an increase of the CN-strong star fraction with cluster 
tidal radius. Since GCs with large tidal radii are mostly found in the weak
tidal field of the Galaxy (well outside the bulge and disk potential) they
might occupy orbits that avoid bulge/disk shocks. Thus they might keep
their gas longer, which favors the build-up of a second generation of enriched 
stars. Furthermore, we find that preferably the more luminous/massive clusters
exhibit a large number of CN-strong stars. This may be an indication that 
the CNO processed ejecta could be more efficiently retained by more massive 
objects, independent of their tidal radius. The picture emerges that there
are two basic channels that lead to a high fraction of CN-rich stars in GCs:
1) the cluster formed and lived in a remote environment, which allowed it to
keep/regain its gas, and 2) the gravitational potential of the cluster itself
was large enough to trap the enriched ejecta of slow velocity winds out of which
a new generation of stars was formed. 

Interestingly, those clusters that do not follow the observed trend are either
associated with the young halo or accreted from the Sgr dSph. This might 
indicate that, as third parameter, the environmental conditions in which the 
clusters formed might had a non-negligible influence on the abundance patterns
we observe today.

Nevertheless we point out that our study is limited to a small sample of 
clusters. For a statistically better supported study a larger cluster sample is 
necessary. Furthermore a complete set of cluster parameters are needed to 
search for the significance of the CN-CH differences between genuine halo globular 
clusters and accreted objects.


\acknowledgements
The authors thank the referee for the useful comments and suggestions. 
A.K. and E.K.G. gratefully acknowledge support by the Swiss National
Science Foundation through the grants 200020-105260 and 200020-113697.
M.H. acknowledges support from a German Science Foundation Grant (DFG-Projekt
HI-855/2).


\appendix
\section{List of spectroscopic sample stars}

The following table contains a magnitude limited list (five stars per cluster)
of our spectroscopic sample stars. They are ordered by increasing $V$ 
magnitude. The columns are as follows:\\

\noindent
{\bf Column 1.} Identification of the object, giving the name of the globular
cluster followed by a number which is ordered with increasing $V$ magnitude.

\noindent
{\bf Column 2.} Right ascension for the epoch 2000 in decimals.

\noindent
{\bf Column 3.} Declination (2000) in decimals.

\noindent
{\bf Column 4.} Apparent $V$ magnitude as determined by PSF photometry with
{\sc DAOPHOT II} under {\sc IRAF}.

\noindent
{\bf Column 5.} $B-V$ colour from PSF photometry.

\noindent
{\bf Column 6.} Measured CN band strength.

\noindent
{\bf Column 7.} Error in measured CN band strength.

\noindent
{\bf Column 8.} Measured CH band strength.

\noindent
{\bf Column 9.} Error in measured CH band strength.

\noindent
{\bf Column 10.} Calculated CN-excess parameter \dcn.

\noindent
{\bf Column 11.} Radial velocity as determined from cross-correlation with
{\sc FXCOR} under {\sc IRAF} and not corrected for systematic errors. Thus,
these velocities are only indicative.

\noindent
{\bf Column 12.} `Type' describes to which part of the CMD the star most 
probably belongs: RGB = red giant branch, HB = horizontal branch, SGB = sub giant branch, MS = main 
sequence.\\

{\bf Note:} The full table of analyzed stars only is available in the online 
version of the article.

\begin{table*}[h]
\centering
\caption{\label{tabAPP} List of spectroscopic sample stars, ordered by 
increasing $V$ magnitude.}
\begin{tabular}{lrrccrrrrrrl}
\hline\hline
Id          & $\alpha$(2000) & $\delta$(2000)  & $V$        & $B-V$   & CN     &dCN      & CH    &dCH     & \dcn     & $v_{rad}$ & Type \\
\hline
NGC288-001 	 & $13.23449$ & $-26.53142$ & $16.83$ & $ 0.76$ & $-0.493$ & $ 0.105$ & $ 2.291$ & $ 0.056$ & $ 0.303$ & $ -90.9$ & RGB\\
NGC288-002 	 & $13.22360$ & $-26.59697$ & $16.97$ & $ 0.78$ & $-0.546$ & $ 0.094$ & $ 2.288$ & $ 0.052$ & $ 0.252$ & $ -91.1$ & RGB\\
NGC288-003 	 & $13.20552$ & $-26.63372$ & $17.14$ & $ 0.76$ & $-0.482$ & $ 0.082$ & $ 2.355$ & $ 0.049$ & $ 0.319$ & $ -83.4$ & RGB\\
NGC288-004 	 & $13.22098$ & $-26.54228$ & $17.37$ & $ 0.71$ & $ 0.265$ & $ 0.130$ & $ 2.125$ & $ 0.063$ & $ 1.070$ & $ -89.1$ & RGB\\
NGC288-005 	 & $13.22482$ & $-26.60748$ & $17.52$ & $ 0.73$ & $-0.004$ & $ 0.120$ & $ 2.124$ & $ 0.063$ & $ 0.803$ & $ -94.3$ & RGB\\
... & ... & ... & ... & ... & ... & ... & ... & ... & ... & ... & ...\\
NGC362-001 	 & $15.70094$ & $-70.86414$ & $16.73$ & $ 0.81$ & $-0.051$ & $ 0.066$ & $ 2.115$ & $ 0.039$ & $ 0.822$ & $ 219.7$ & RGB\\
\hline\hline
\end{tabular}
\end{table*}


\bibliographystyle{aa}
\bibliography{9446}

\begin{thebibliography}{69}
\expandafter\ifx\csname natexlab\endcsname\relax\def\natexlab#1{#1}\fi

\bibitem[{{Alcaino} {et~al.}(1997){Alcaino}, {Liller}, \& {Alvarado}}]{A/L:97}
{Alcaino}, G., {Liller}, W., \& {Alvarado}, F. 1997, \aj, 114, 2626

\bibitem[{{Alcaino} {et~al.}(1992){Alcaino}, {Liller}, {Alvarado}, \&
  {Wenderoth}}]{A/L:92}
{Alcaino}, G., {Liller}, W., {Alvarado}, F., \& {Wenderoth}, E. 1992, \aj, 104,
  190

\bibitem[{{Bedin} {et~al.}(2004){Bedin}, {Piotto}, {Anderson}, {Cassisi},
  {King}, {Momany}, \& {Carraro}}]{B/P:04}
{Bedin}, L.~R., {Piotto}, G., {Anderson}, J., {et~al.} 2004, \apjl, 605, L125

\bibitem[{{Bekki} {et~al.}(2007){Bekki}, {Campbell}, {Lattanzio}, \&
  {Norris}}]{B/C:07}
{Bekki}, K., {Campbell}, S.~W., {Lattanzio}, J.~C., \& {Norris}, J.~E. 2007,
  \mnras, 377, 335

\bibitem[{{Bellazzini} {et~al.}(2003){Bellazzini}, {Ferraro}, \&
  {Ibata}}]{B/F/I:03}
{Bellazzini}, M., {Ferraro}, F.~R., \& {Ibata}, R. 2003, \aj, 125, 188

\bibitem[{{Bellazzini} {et~al.}(2001){Bellazzini}, {Pecci}, {Ferraro},
  {Galleti}, {Catelan}, \& {Landsman}}]{B/P:01}
{Bellazzini}, M., {Pecci}, F.~F., {Ferraro}, F.~R., {et~al.} 2001, \aj, 122,
  2569

\bibitem[{{Briley}(1997)}]{B:97}
{Briley}, M.~M. 1997, \aj, 114, 1051

\bibitem[{{Briley} {et~al.}(1989){Briley}, {Bell}, {Smith}, \& {Hesser}}]{B:89}
{Briley}, M.~M., {Bell}, R.~A., {Smith}, G.~H., \& {Hesser}, J.~E. 1989, \apj,
  341, 800

\bibitem[{{Briley} {et~al.}(2004){Briley}, {Harbeck}, {Smith}, \&
  {Grebel}}]{B/H:04}
{Briley}, M.~M., {Harbeck}, D., {Smith}, G.~H., \& {Grebel}, E.~K. 2004, \aj,
  127, 1588

\bibitem[{{Briley} {et~al.}(2007){Briley}, {Martell}, \& {Smith}}]{B/M/S:07}
{Briley}, M.~M., {Martell}, S., \& {Smith}, G.~H. 2007, in American
  Astronomical Society Meeting Abstracts, Vol. 211, American Astronomical
  Society Meeting Abstracts, 31--

\bibitem[{{Buonanno} {et~al.}(1998){Buonanno}, {Corsi}, {Pulone}, {Fusi Pecci},
  \& {Bellazzini}}]{B/C:98}
{Buonanno}, R., {Corsi}, C.~E., {Pulone}, L., {Fusi Pecci}, F., \&
  {Bellazzini}, M. 1998, \aap, 333, 505

\bibitem[{{Buonanno} {et~al.}(1995){Buonanno}, {Corsi}, {Pulone}, {Pecci},
  {Richer}, \& {Fahlman}}]{B/C:95}
{Buonanno}, R., {Corsi}, C.~E., {Pulone}, L., {et~al.} 1995, \aj, 109, 663

\bibitem[{{Carretta}(2006)}]{C:06}
{Carretta}, E. 2006, \aj, 131, 1766

\bibitem[{{Charbonnel}(1995)}]{C:95}
{Charbonnel}, C. 1995, \apjl, 453, L41

\bibitem[{{Cohen}(1978)}]{C:78}
{Cohen}, J.~G. 1978, \apj, 223, 487

\bibitem[{{Cohen}(1999)}]{C:99.2}
---. 1999, \aj, 117, 2434

\bibitem[{{Cohen}(2004)}]{C:04}
---. 2004, \aj, 127, 1545

\bibitem[{{Cottrell} \& {Da Costa}(1981)}]{C/DC:81}
{Cottrell}, P.~L. \& {Da Costa}, G.~S. 1981, \apjl, 245, L79

\bibitem[{{D'Antona} {et~al.}(2005){D'Antona}, {Bellazzini}, {Caloi}, {Pecci},
  {Galleti}, \& {Rood}}]{DA/B:05}
{D'Antona}, F., {Bellazzini}, M., {Caloi}, V., {et~al.} 2005, \apj, 631, 868

\bibitem[{{D'Antona} {et~al.}(2002){D'Antona}, {Caloi}, {Montalb{\'a}n},
  {Ventura}, \& {Gratton}}]{DA/C:02}
{D'Antona}, F., {Caloi}, V., {Montalb{\'a}n}, J., {Ventura}, P., \& {Gratton},
  R. 2002, \aap, 395, 69

\bibitem[{{D'Antona} {et~al.}(1983){D'Antona}, {Gratton}, \&
  {Chieffi}}]{DA/G:83}
{D'Antona}, F., {Gratton}, R., \& {Chieffi}, A. 1983, Mem.SAIt, 54, 173

\bibitem[{{Decressin} {et~al.}(2007){Decressin}, {Meynet}, {Charbonnel},
  {Prantzos}, \& {Ekstr{\"o}m}}]{D/M:07}
{Decressin}, T., {Meynet}, G., {Charbonnel}, C., {Prantzos}, N., \&
  {Ekstr{\"o}m}, S. 2007, \aap, 464, 1029

\bibitem[{{Denissenkov} \& {Herwig}(2003)}]{D/H:03}
{Denissenkov}, P.~A. \& {Herwig}, F. 2003, \apjl, 590, L99

\bibitem[{{Denissenkov} \& {VandenBerg}(2003)}]{D/VdB:03}
{Denissenkov}, P.~A. \& {VandenBerg}, D.~A. 2003, \apj, 593, 509

\bibitem[{{Durrell} \& {Harris}(1993)}]{D/H:93}
{Durrell}, P.~R. \& {Harris}, W.~E. 1993, \aj, 105, 1420

\bibitem[{{Frenk} \& {Fall}(1982)}]{F/F:82}
{Frenk}, C.~S. \& {Fall}, S.~M. 1982, \mnras, 199, 565

\bibitem[{{Fulbright}(2002)}]{F:02}
{Fulbright}, J.~P. 2002, \aj, 123, 404

\bibitem[{{Gratton} {et~al.}(2004){Gratton}, {Sneden}, \&
  {Carretta}}]{G/S/C:04}
{Gratton}, R., {Sneden}, C., \& {Carretta}, E. 2004, \araa, 42, 385

\bibitem[{{Gratton} {et~al.}(2000){Gratton}, {Sneden}, {Carretta}, \&
  {Bragaglia}}]{G/S:00}
{Gratton}, R.~G., {Sneden}, C., {Carretta}, E., \& {Bragaglia}, A. 2000, \aap,
  354, 169

\bibitem[{{Harbeck} {et~al.}(2003{\natexlab{a}}){Harbeck}, {Smith}, \&
  {Grebel}}]{H/S/G:03}
{Harbeck}, D., {Smith}, G.~H., \& {Grebel}, E.~K. 2003{\natexlab{a}}, \aj, 125,
  197

\bibitem[{{Harbeck} {et~al.}(2003{\natexlab{b}}){Harbeck}, {Smith}, \&
  {Grebel}}]{H/S/G:03b}
---. 2003{\natexlab{b}}, \aap, 409, 553

\bibitem[{{Harris}(1996)}]{H:96}
{Harris}, W.~E. 1996, \aj, 112, 1487

\bibitem[{{Hilker} {et~al.}(2004){Hilker}, {Kayser}, {Richtler}, \&
  {Willemsen}}]{H/K:04}
{Hilker}, M., {Kayser}, A., {Richtler}, T., \& {Willemsen}, P. 2004, \aap, 422,
  L9

\bibitem[{{Iben}(1968)}]{I:68}
{Iben}, Jr., I. 1968, \nat, 220, 143

\bibitem[{{Kayser} {et~al.}(2006){Kayser}, {Hilker}, {Richtler}, \&
  {Willemsen}}]{K/H:06}
{Kayser}, A., {Hilker}, M., {Richtler}, T., \& {Willemsen}, P.~G. 2006, \aap,
  458, 777

\bibitem[{{Kraft}(1994)}]{K:94}
{Kraft}, R.~P. 1994, \pasp, 106, 553

\bibitem[{{Lee}(2000)}]{L:00}
{Lee}, S.~G. 2000, Journal of Korean Astronomical Society, 33, 137

\bibitem[{{Lee}(2005)}]{L:05}
---. 2005, Journal of Korean Astronomical Society, 38, 23

\bibitem[{{Mackey} \& {van den Bergh}(2005)}]{M/VDB:05}
{Mackey}, A.~D. \& {van den Bergh}, S. 2005, \mnras, 360, 631

\bibitem[{{Maeder} \& {Meynet}(2006)}]{M/M:06}
{Maeder}, A. \& {Meynet}, G. 2006, \aap, 448, L37

\bibitem[{{Norris}(1987)}]{N:87}
{Norris}, J. 1987, \apjl, 313, L65

\bibitem[{{Norris} \& {Smith}(1981)}]{N/S:81}
{Norris}, J. \& {Smith}, G.~H. 1981, in IAU Colloq. 68: Astrophysical
  Parameters for Globular Clusters, ed. A.~G.~D. {Philip} \& D.~S. {Hayes}, 109

\bibitem[{{Osborn}(1971)}]{O:71}
{Osborn}, W. 1971, The Observatory, 91, 223

\bibitem[{{Penny} {et~al.}(1992){Penny}, {Smith}, \& {Churchill}}]{P:92}
{Penny}, A.~J., {Smith}, G.~H., \& {Churchill}, C.~W. 1992, \mnras, 257, 89

\bibitem[{{Piotto} {et~al.}(2007){Piotto}, {Bedin}, {Anderson}, {King},
  {Cassisi}, {Milone}, {Villanova}, {Pietrinferni}, \& {Renzini}}]{P/B:07}
{Piotto}, G., {Bedin}, L.~R., {Anderson}, J., {et~al.} 2007, \apjl, 661, L53

\bibitem[{{Pritzl} {et~al.}(2005){Pritzl}, {Venn}, \& {Irwin}}]{P/V:05}
{Pritzl}, B.~J., {Venn}, K.~A., \& {Irwin}, M. 2005, \aj, 130, 2140

\bibitem[{{Pryor} \& {Meylan}(1993)}]{P/M:93}
{Pryor}, C. \& {Meylan}, G. 1993, in ASP Conf. Ser. 50: Structure and Dynamics
  of Globular Clusters, 357

\bibitem[{{Richter} {et~al.}(1999){Richter}, {Hilker}, \& {Richtler}}]{R/H:99}
{Richter}, P., {Hilker}, M., \& {Richtler}, T. 1999, \aap, 350, 476

\bibitem[{{Rosenberg} {et~al.}(1999){Rosenberg}, {Saviane}, {Piotto}, \&
  {Aparicio}}]{R/S:99}
{Rosenberg}, A., {Saviane}, I., {Piotto}, G., \& {Aparicio}, A. 1999, \aj, 118,
  2306

\bibitem[{{Samus} {et~al.}(1995{\natexlab{a}}){Samus}, {Ipatov}, {Smirnov},
  {Kravtsov}, {Alcaino}, {Liller}, \& {Alvarado}}]{S/I:95}
{Samus}, N., {Ipatov}, A., {Smirnov}, O., {et~al.} 1995{\natexlab{a}}, \aaps,
  112, 439

\bibitem[{{Samus} {et~al.}(1995{\natexlab{b}}){Samus}, {Kravtsov}, {Pavlov},
  {Alcaino}, \& {Liller}}]{S/K:95}
{Samus}, N., {Kravtsov}, V., {Pavlov}, M., {Alcaino}, G., \& {Liller}, W.
  1995{\natexlab{b}}, \aaps, 109, 487

\bibitem[{{Sbordone} {et~al.}(2007){Sbordone}, {Bonifacio}, {Buonanno},
  {Marconi}, {Monaco}, \& {Zaggia}}]{S/B:07}
{Sbordone}, L., {Bonifacio}, P., {Buonanno}, R., {et~al.} 2007, \aap, 465, 815

\bibitem[{{Sbordone} {et~al.}(2005){Sbordone}, {Bonifacio}, {Marconi},
  {Buonanno}, \& {Zaggia}}]{S/B:05}
{Sbordone}, L., {Bonifacio}, P., {Marconi}, G., {Buonanno}, R., \& {Zaggia}, S.
  2005, \aap, 437, 905

\bibitem[{{Shetrone} {et~al.}(2001){Shetrone}, {C{\^o}t{\'e}}, \&
  {Sargent}}]{S/C:01}
{Shetrone}, M.~D., {C{\^o}t{\'e}}, P., \& {Sargent}, W.~L.~W. 2001, \apj, 548,
  592

\bibitem[{{Smith}(2002)}]{S:02}
{Smith}, G.~H. 2002, \pasp, 114, 1215

\bibitem[{{Smith} \& {Martell}(2003)}]{S/M:03}
{Smith}, G.~H. \& {Martell}, S.~L. 2003, \pasp, 115, 1211

\bibitem[{{Smith} \& {Mateo}(1990)}]{S/M:90}
{Smith}, G.~H. \& {Mateo}, M. 1990, \apj, 353, 533

\bibitem[{{Smith} \& {Norris}(1982)}]{S/N:82}
{Smith}, G.~H. \& {Norris}, J. 1982, \apj, 254, 149

\bibitem[{{Smith} \& {Norris}(1983)}]{S/N:83}
---. 1983, \apj, 264, 215

\bibitem[{{Smith} {et~al.}(1996){Smith}, {Shetrone}, {Bell}, {Churchill}, \&
  {Briley}}]{S/S:96}
{Smith}, G.~H., {Shetrone}, M.~D., {Bell}, R.~A., {Churchill}, C.~W., \&
  {Briley}, M.~M. 1996, \aj, 112, 1511

\bibitem[{{Stetson} {et~al.}(1989){Stetson}, {Hesser}, {Smith}, {Vandenberg},
  \& {Bolte}}]{S/H:89}
{Stetson}, P.~B., {Hesser}, J.~E., {Smith}, G.~H., {Vandenberg}, D.~A., \&
  {Bolte}, M. 1989, \aj, 97, 1360

\bibitem[{{Suntzeff}(1981)}]{S:81}
{Suntzeff}, N.~B. 1981, \apjs, 47, 1

\bibitem[{{Sweigart} \& {Mengel}(1979)}]{S/M:79}
{Sweigart}, A.~V. \& {Mengel}, J.~G. 1979, \apj, 229, 624

\bibitem[{{Thoul} {et~al.}(2002){Thoul}, {Jorissen}, {Goriely}, {Jehin},
  {Magain}, {Noels}, \& {Parmentier}}]{T/J:02}
{Thoul}, A., {Jorissen}, A., {Goriely}, S., {et~al.} 2002, \aap, 383, 491

\bibitem[{{Ventura} {et~al.}(2001){Ventura}, {D'Antona}, {Mazzitelli}, \&
  {Gratton}}]{V/DA:01}
{Ventura}, P., {D'Antona}, F., {Mazzitelli}, I., \& {Gratton}, R. 2001, \apjl,
  550, L65

\bibitem[{{Weiss} {et~al.}(2000){Weiss}, {Denissenkov}, \&
  {Charbonnel}}]{W/D:00}
{Weiss}, A., {Denissenkov}, P.~A., \& {Charbonnel}, C. 2000, \aap, 356

\bibitem[{{Willemsen} {et~al.}(2005){Willemsen}, {Hilker}, {Kayser}, \&
  {Bailer-Jones}}]{W/H:05}
{Willemsen}, P.~G., {Hilker}, M., {Kayser}, A., \& {Bailer-Jones}, C.~A.~L.
  2005, \aap, 436, 379

\bibitem[{{Zinn}(1985)}]{Z:85}
{Zinn}, R. 1985, \apj, 293, 424

\bibitem[{{Zinn}(1993)}]{Z:93}
{Zinn}, R. 1993, in ASP Conf. Ser., Vol.~48, The Globular Cluster-Galaxy
  Connection, ed. G.~H. {Smith} \& J.~P. {Brodie}, 38

\end{thebibliography}


\Online

\longtab{1}{
\begin{longtable}{lrrccrrrrrrl}
\caption{\label{tabOnline} List of spectroscopic sample stars, ordered by increasing $V$ magnitude.}\\
\hline\hline
Id          & $\alpha$(2000) & $\delta$(2000)  & $V$  & $B-V$  & CN   &dCN   & CH   &dCH  & \dcn   & $v_{rad}$ & Type \\
\hline
\endfirsthead
\caption{continued.}\\
\hline\hline
Id          & $\alpha$(2000) & $\delta$(2000)  & $V$  & $B-V$  & CN   &dCN   & CH   &dCH  & \dcn   & $v_{rad}$ & Type \\
\hline
\endhead
\hline
\endfoot
NGC288-001 	 & $13.23449$ & $-26.53142$ & $16.83$ & $ 0.76$ & $-0.493$ & $ 0.105$ & $ 2.291$ & $ 0.056$ & $ 0.303$ & $ -90.9$ & RGB\\
NGC288-002 	 & $13.22360$ & $-26.59697$ & $16.97$ & $ 0.78$ & $-0.546$ & $ 0.094$ & $ 2.288$ & $ 0.052$ & $ 0.252$ & $ -91.1$ & RGB\\
NGC288-003 	 & $13.20552$ & $-26.63372$ & $17.14$ & $ 0.76$ & $-0.482$ & $ 0.082$ & $ 2.355$ & $ 0.049$ & $ 0.319$ & $ -83.4$ & RGB\\
NGC288-004 	 & $13.22098$ & $-26.54228$ & $17.37$ & $ 0.71$ & $ 0.265$ & $ 0.130$ & $ 2.125$ & $ 0.063$ & $ 1.070$ & $ -89.1$ & RGB\\
NGC288-005 	 & $13.22482$ & $-26.60748$ & $17.52$ & $ 0.73$ & $-0.004$ & $ 0.120$ & $ 2.124$ & $ 0.063$ & $ 0.803$ & $ -94.3$ & RGB\\
NGC288-006 	 & $13.23787$ & $-26.52769$ & $17.57$ & $ 0.70$ & $-0.034$ & $ 0.128$ & $ 2.158$ & $ 0.066$ & $ 0.774$ & $ -94.7$ & RGB\\
NGC288-007 	 & $13.23132$ & $-26.57133$ & $17.62$ & $ 0.72$ & $ 0.121$ & $ 0.110$ & $ 2.176$ & $ 0.060$ & $ 0.930$ & $ -79.7$ & RGB\\
NGC288-008 	 & $13.24678$ & $-26.52963$ & $17.71$ & $ 0.69$ & $-0.034$ & $ 0.137$ & $ 2.133$ & $ 0.071$ & $ 0.776$ & $ -91.9$ & RGB\\
\end{longtable}
}


\end{document}